\documentclass[prx,twocolumn,superscriptaddress,nofootinbib,showpacs,notitlepage]{revtex4-2}
\usepackage{amsmath}
\usepackage{amsfonts}
\usepackage{amssymb}
\usepackage{bbm}
\usepackage[utf8]{inputenc}
\usepackage{braket}
\usepackage{environ}
\usepackage{graphicx,color,dsfont}
\usepackage{booktabs}
\usepackage{floatrow}
\usepackage[caption=false,label font={bf,normalsize}]{subfig}
\usepackage[dvipsnames]{xcolor}
\usepackage[all]{xy}
\usepackage{relsize}
\usepackage{hyperref}
\usepackage{accents}
\hypersetup{linktocpage}
\definecolor{darkblue}{RGB}{0,0,127} 
\definecolor{darkgreen}{RGB}{0,150,0}
\hypersetup{colorlinks=true,citecolor=darkblue,linkcolor=darkblue, 
	filecolor=red, urlcolor=darkblue,pdftitle={Classifying  
Logical Gates in Quantum Codes via Cohomology Operations and Symmetry},
pdfauthor={Po-Shen Hsin, Ryohei Kobayashi, Guanyu Zhu}}
\usepackage{lmodern}

\usepackage{lipsum}

\makeatletter
\def\@opargbegintheorem#1#2#3{\trivlist
   \item[]{\bfseries #1\ #2\ (#3)} \itshape}
\makeatother

\NewEnviron{eqs}{%
\begin{equation}\begin{split}
    \BODY
\end{split}\end{equation}
}

\newcommand{\ZZ}{\mathbb{Z}}
\newcommand{\Z}{\mathbb{Z}}
\newcommand{\M}{\mathcal{M}}

\begin{document}

\title{Classifying  
Logical Gates 
 in Quantum  Codes
via Cohomology Operations and Symmetry}

\author{Po-Shen Hsin}
\email{po-shen.hsin@kcl.ac.uk}

\affiliation{Department of Mathematics, King’s College London, Strand, London WC2R 2LS, UK}

\author{Ryohei Kobayashi}
\email{ryok@ias.edu}

\affiliation{School of Natural Sciences, Institute for Advanced Study, Princeton, NJ 08540, USA}

\author{Guanyu Zhu}
\email{guanyu.zhu@ibm.com}
\affiliation{IBM Quantum, IBM T.J. Watson Research Center, Yorktown Heights, NY 10598 USA}

\begin{abstract}

We systematically construct and classify fault-tolerant logical gates implemented by constant-depth circuits for quantum codes using cohomology operations and symmetry. These logical gates are obtained from unitary operators given by symmetry-protected topological responses, which correspond to generators of group cohomology and can be expressed explicitly on the lattice using cohomology operations including cup product, Steenrod squares and new combinations of higher cup products called higher Pontryagin powers.  Our study covers most types of the cohomology operations in the literature. This hence gives rise to logical $C^{n-1}Z$ gates in $n$ copies of quantum codes via the $n$-fold cup product in the usual color code paradigm, as well as several new classes of diagonal and non-diagonal logical gates in increasing Clifford hierarchies beyond the color code paradigm, including the logical $R_k$ and multi-controlled $C^m R_k$ gates for codes defined in projective spaces. Implementing these gates could make it more efficient to compile specific types of quantum algorithms such as Shor's algorithm.   We further extend the construction to quantum codes with boundaries,  which generalizes the folding approach in color codes. We also present a formalism for addressable and parallelizable logical gates in LDPC codes via higher-form symmetries.  We further construct logical Clifford gates in expander-based codes including the asymptotically good LDPC codes and hypergraph-product codes. As a byproduct, we find new topological responses of finite higher-form symmetries using higher Pontryagin~powers.

\end{abstract}

\maketitle
\tableofcontents

\unitlength = .8mm

\setcounter{tocdepth}{3}

\section{Introduction}

A fundamental problem in fault-tolerant quantum computing is to find the optimal  spacetime overhead required for universal computation.  In recent years,  there has been significant progress on quantum low-density parity check (qLDPC) codes \cite{fiberbundlecode21,9567703,9490244,hastingswr21,pkldpc22,Quantum_tanner, lh22,guefficient22,dhlv23,lzdecoding23,gusingleshot23, Bravyi:2024wc}.  In particular, a family of asymptotically good qLDPC codes have been found which achieves constant overhead and linear distance for a quantum memory \cite{pkldpc22, Quantum_tanner}.  However, it remains an open question what would be the optimal fault-tolerant quantum processor.   In particular, many of the proposed qLDPC codes with small overhead may not supply powerful logical gates such as non-Clifford gates.   Therefore, systematic study and classification of possible logical gates in quantum codes is a crucial task for the advance of fault-tolerant computation.

By far, most of the previous studies of transversal or constant-depth logical gates in quantum codes are based on ad hoc combinatorial methods \cite{Bombin:2007eh, bravyi2012magic,  Bombin_2015, Kubica2015, Kubica:2015mta, bombin2018transversal, bombin20182d, haah2018codes,  vasmer2019fault, JochymOConnor:2021ih, Vasmer:2022morphing, Vuillot2022, wills2024constant, nguyen2024good, golowich2024asymptotically, scruby2024quantum}. As for qLDPC codes, most constructions of transversal logical non-Clifford gates (or more generally non-Clifford gates based on constant-depth circuits) can be understood as unfolding of color codes \cite{Kubica:2015mta}, which we call the \textit{color code paradigm}.  

It has been recently realized that the transversal non-Clifford gate in color codes can be understood as a \textit{cohomology operation} with cup products \cite{zhu2023non}, based on  previous understanding of the gauged symmetry-protected topological (SPT) defects and the corresponding higher symmetries in a $\ZZ_2^3$ gauge theory  \cite{barkeshli2023codimension, Barkeshli:2022edm}.  It was further realized in Ref.~\cite{zhu2023non} that applying such a formalism leads to the first non-Clifford logical gate on a high-rate qLDPC codes defined on 3-manifolds. This opens up the possibility to systematically classifying logical gates based on cohomology operation and more generally emergent symmetries \cite{Gaiotto_2015, Yoshida_gate_SPT_2015, Yoshida_global_symmetry_2016, Yoshida2017387, Zhu:2017tr, barkeshli2023codimension, Barkeshli:2022edm, zhu:2022fractal, song2024magic} in general quantum codes.  In particular, it is desirable to go beyond the color code paradigm, which is equivalent to a $n$-fold cup product of $n$ different cocycles according to Ref.~\cite{zhu2023non}. 

From the point of view of algebraic topology, quantum CSS codes and qLDPC codes are all based on chain complexes.  It is then natural to expect that a large class of logical gates should be based on the cohomology operations on the associated chain complexes.   The simplest type of such is the cup product operation, which is what color code logical gates are based on. Meanwhile, there exist many other types of cohomology operations, such as Pontryagin powers, Steenrod squares and higher cup products, which can give rise to many more interesting logical gates.

From the physics perspective, the classification of the large class of logical gates from cohomology operation is to some extent equivalent to classifying SPT phases, their corresponding group cohomology classes and the associated SPT actions \cite{Yoshida_gate_SPT_2015, Yoshida_global_symmetry_2016, Yoshida2017387, barkeshli2023codimension, Barkeshli:2022edm, zhu2023non}.   The extra subtlety in classifying logical gates is that it is a combination of the type of cohomology operation and the topology of the underlying chain complex such as various types of manifolds in the context of homological codes, as well as the corresponding  boundary conditions.  It is expected to be a gigantic program to completely classify all possible logical gates corresponding to cohomology operations.   In this paper, we instead develop a general formalism that can be used for this classification which touches the majority types of cohomology operations in the literature, and also identify certain interesting families of new logical gates that arise from this classification.

The mathematical formalism we have developed is based on \textit{operator-valued cochains} using the algebraic topology language.  Physically, this is also called a \textit{gauge field formalism}, since it originates from the underlying topological quantum field theory (TQFT) or equivalently the  lattice gauge theory corresponding to the quantum code. It has also been shown that general quantum CSS and LDPC codes can be related to gauging  \cite{kubica2018ungauging, rakovszky2023physicsgoodldpccodes,rakovszky2024physicsgoodldpccodes}.  The operator-valued cochains represent the gauge fields in this formalism, and can also be related to Pauli operators, stabilizers and logical operators in the quantum  code.      We further use these operator-valued cochains to define cohomology operations such as the cup products on a simplicial or cubical complex, or even more generally a cell complex. We use this formalism to construct logical gates in arbitrary Clifford hierarchies based on cup products on generic simplicial or cubical complexes, including the case of homological codes defined on the triangulation or cellulation of a manifold. The formalism also gives the explicit constant-depth local circuits implementing these logical gates. We also construct addressable and parallelizable logical gates using higher-form ($q$-form) symmetries which only act on a codimension-$k$ subcomplex or submanifold. 
We further generalize our formalism to higher-form gauge theory, which describes higher-form homological codes where the qubits are placed on $i$-cells with $i \ge 2$. This type of theory is relevant for self-correcting quantum memories, as well as qLDPC codes based on high-dimensional expanders, such as the code in  Ref.~\cite{Guth:2014cj}.   

Although the majority examples we consider in this paper are homological codes, we also extend this formalism to construct logical Clifford gates in expander-based qLDPC codes such as the asymptotically good qLDPC codes \cite{pkldpc22, Quantum_tanner} and the hypergraph product code \cite{Tillich:2014_hyergraph_product} using the code-to-manifold mapping introduced in Ref.~\cite{freedman:2020_manifold_from_code}.  We leave the more general extension of this formalism to general qLDPC or CSS codes with logical non-Clifford gates to Ref.~\cite{zhu2025topological}. 

In order to go beyond the color code paradigm ($n$-fold cup product), we consider more general classes of cohomology operations that cover most of the operations known in the literature (e.g. \cite{mosher2008cohomology}), such as higher Pontryagin powers and Steenrod squares, that construct effective action of symmetry-protected topological (SPT) phases on subspaces. 
 In particular, we will focus on the logical gates in higher-form homological codes and the corresponding higher-form gauge theories. For $n$-form homological code describing $G$ gauge theory in the ground states, we will focus on the logical gates correspond to the group cohomology $H^*(B^nG,U(1))$.\footnote{
There are also beyond group cohomology SPT phases \cite{Kapustin:2014tfa}. Here, we focus on group cohomology phases, since they have direct realization on physical qubits using cup product and higher cup products.
}  In particular, the higher Pontryagin powers lead to several family of new logical gates with increasing Clifford hierarchy.   This includes the family of logical $R_k$ gate [$R_k=\mathrm{diag}\left(1,\exp({2\pi i}/{2^{k}})\right)$] for homological codes defined on projective manifolds including $\mathbb{CP}^{n/2}$ and $\mathbb{RP}^{n}$.  This is surprising since such logical gates only exist for color codes defined on open manifolds with gapped boundaries, while the new logical gates we discover exist in even closed manifold.   These new gates would be more useful in the context of high-rate qLDPC codes, since such codes are typically defined on complexes or manifolds without boundaries.  

Even more interestingly, the higher Pontryagin powers also give rise to new multi-controlled logical $C^m R_k$ gates, which has not been discovered in previous quantum codes.   For example, the logical $C R_k$ gates appear frequently in certain quantum algorithm such as the quantum Fourier transform, which is an important subroutine of the Shor's algorithm, as well as quantum phase estimation and quantum simulation algorithms \cite{nielsen_chuang_2010}, and even certain quantum machine learning algorithms \cite{Lloyd:2014gcb}.  A direct implementation of such logical gates fault-tolerantly can significantly improve the logical circuit compiling.   This opens up a new direction of searching algorithm-efficient logical gates in quantum codes.  

While the previous studies of cohomology operation as logical gates were all based on closed complexes or manifolds \cite{zhu2023non}, we also generalize such descriptions to codes in the presence of boundaries in this paper. From the mathematical perspective, this corresponds to the study of relative homology/cohomology for a chain complex.   In particular, we generalize the folding approach in color codes by defining boundary operation ${\cal B}$ that constructs well-defined gapped boundaries of SPT phases without introducing new physical qubits. The boundaries can be codimension one, codimension two, etc, such as hinges or corners, and the boundary operators are obtained from the bulk SPT phase using the boundary map ${\cal B}$. Such operator with the boundary gives fault-tolerant logical gates directly applied to copies of  homological codes and also gives a topological interpretation of the corresponding transversal logical gates in the color codes \cite{Kubica:2015mta}.  Physically, the symmetries considered here are gauged SPT symmetries \cite{Barkeshli:2022edm,Barkeshli:2023bta}, and their symmetry defects are condensation defects and therefore admit gapped boundaries-- which provides the boundary description used in this approach.

Together with the invertible electromagnetic duality for untwisted Abelian $n$-form homological codes in $2n$ space dimension, which gives Hadamard logical gate, and automorphism symmetry of the gauge group that gives suitable SWAP logical gates, the above formalism provides a classification of the fault-tolerant logical gates in homological codes.

The work is organized as follows. In Sec.~\ref{sec:reviewgaugefield}, we introduce the gauge field formalism for quantum codes and define the cup products on simplicial or cubical complexes, as well as the use of cup product to obtain logical $C^{n-1}Z$ gate. We also discuss addressable and parallelizable logical gates via higher form symmetries and the generalized formalism for higher-form gauge theory. We further study the non-diagonal logical gates such as CNOTs in the form of cohomology operation. In Sec.~\ref{sec:higherPontryagin}, we introduce higher Pontryagin power cohomology operation in SPT phase and obtain new fault-tolerant non-Clifford logical gates in higher-form homological codes. In Sec.~\ref{sec:logicalgatebdySPT}, we generalize the folding construction and define boundary operation of SPT phases, and construct fault-tolerant non-Clifford logical gates from boundary of SPT phases. In Sec.~\ref{sec:moreexample}, we discuss more cohomology operation given by Steenrod squares and discuss their application in logical gates from SPT phases and boundary operation for higher-form SPT phases. In Sec.~\ref{sec:goodqLDPC}, we construct  logical Clifford gates via cohomology operations for good quantum LDPC codes or hypergraph product codes.
We summarize the work and discuss future directions in Sec.~\ref{sec:discussion}.
There are three appendices. In appendix \ref{sec:cupproductrev}, we review the definition of cup products on triangulated lattices and hypercubic lattices. In appendix \ref{sec:higherP}, we give the explicit expression of higher Pontryagin power on any lattice using cup product and Steenrod's higher cup products, and show it is a cohomology operation. in appendix \ref{sec:condesationZN}, we review the presentation of $\mathbb{Z}_N$ homological code from condensing suitable electric charge in $\mathbb{Z}_{N^2}$ homological code.

\textit{Note ---} During the preparation of this manuscript, we became aware of several other works on logical gates with cup products \cite{breuckmann2024cups, lin2024transversal, golowich2024quantum}. While these works focus on the study of the $n$-fold cup product within the color code paradigm as in Ref.~\cite{zhu2023non} and try to generalize it to expander-based qLDPC codes, our work focuses on the  complete classification of all types of cohomology operations which go beyond the color code paradigm, although we also generalize some of the CLifford gates to good qLDPC codes and hypergraph-product codes. The study of logical non-Clifford gates in qLDPC codes is contained in Ref.~\cite{zhu2025topological}.

\section{Gauge Field Formalism for Logical Gates in Stabilizer Codes defined on simplicial and cubical complexes}
\label{sec:reviewgaugefield}

Let us review the operator-valued cochain formalism (physically corresponding to a gauge field formalism) for stabilizer codes that describe gauge theories (see e.g. \cite{Hsin:2021mjn,Chen:2021xks,Chen:2021xuc,Barkeshli:2022wuz,Barkeshli:2022edm,Hsin:2023ooo,Barkeshli:2023bta, zhu2023non, Hsin:2024pdi}).
There are several advantages:
\begin{itemize}
    \item The formalism naturally fits to the language of algebraic topology.  In particular, it translates the Pauli operators, stabilizers and logical operators to operator-valued cochains.

    \item The formalism allows simple derivation and unification of gate identities as we will show in Sec.~\ref{sec:commutationCnZ} for $C^{n-1}Z$ logical gates in copies of homological code -- they become the field theory statement of certain topological action being gauge invariant.
    
    \item The formalism makes physical meaning of logical gates more transparent--this allows input from physics in easier ways.
\end{itemize}

We will illustrate the formalism using $\mathbb{Z}_2$ 1-form gauge theory on $n$-dimensional space, with qubit on edges of the spatial lattice. In Sec.~\ref{sec:ZNhigherform} we will discuss the straightforward generalization to $\mathbb{Z}_N$ $q$-form gauge theory, with $\mathbb{Z}_N$ qudits on $q$-simplices of the spatial lattice.

\subsection{Gauge fields and dual gauge fields}

The $\mathbb{Z}_2$ gauge field corresponds to an operator-valued cochain $a$ with eigenvalue $0,1$ on every edge $e$ of a simplicial complex $\mathcal{L}$ (or more generally a cell complex). It is related to Pauli $Z_e$ operator on edge $e$ as
\begin{equation}
    Z_e=(-1)^{a(e)}~.
\end{equation}
The logical-$Z$ operator corresponds to Wilson line electric string operator as
\begin{equation}
    W(\gamma)=(-1)^{\int_\gamma a}=(-1)^{\sum_{e\in\gamma} a(e) }=\prod _{e\in \gamma} Z_{e} \equiv \overline{Z}(\gamma)~.
\end{equation}
Here, $\gamma=\sum_{e\in\gamma} e$ is a non-trivial 1-cycle where the operator is supported.
The dual gauge field $b$ on the dual complex $\mathcal{L}^*$ acts on $(d-1)$-cell $s$, is given by\footnote{
In field theory, the gauge theory can be described by the BF theory action $\frac{2}{2\pi}\int adb$~.
}
\begin{equation}
    X_{s^\vee}=(-1)^{b(s)}~,
\end{equation}
where $s^\vee$ is the edge on the original complex $\mathcal{L}$ that intersects $s$. 
The logical-$X$ operator corresponds to the magnetic operator 
\begin{equation}
    M(\Sigma)=(-1)^{\int_{\Sigma} b}=\prod _{e\cap \Sigma\neq 0} X_{e} \equiv \overline{X}(\Sigma)~,
\end{equation}
where $\Sigma$ is a non-trivial $(d-1)$-cycle of the dual complex $\mathcal{L}^*$ cutting the edges $e$.
The Pauli commutation relation matches with the commutation relation of $a,b$-- this is the Aharonov-Bohm phase in $\mathbb{Z}_2$ gauge theory. 

The electric string operator and the magnetic operator has the commutation relation
\begin{align}
    W(\gamma)M(\Sigma) = M(\Sigma)W(\gamma) (-1)^{\sharp(\gamma,\Sigma)}~,
\end{align}
where $\sharp(\gamma,\Sigma)$ is the mod 2 intersection number between $\gamma$ and $\Sigma$. This allows us to identify $W(\gamma), Z(\Sigma)$ with nontrivial intersection as the logical Pauli $Z,X$ operators.

The Gauss law operator for $a$ corresponds to the $X$-stabilizer on the vertex $v$
\begin{equation}
G_v=\prod_{e:\partial e\supset v} X_e=(-1)^{\int_{v^\vee} db}~,   
\end{equation}
where $v^\vee$ is the $d$-dimensional face on the dual complex $\mathcal{L}^*$ that intersects vertex $v$ of the original complex at a single point. This is the divergence of the electric field for the $\mathbb{Z}_2$ gauge field.

\subsection{Commutation relations}

The Pauli commutation relations can be phrased as follows:
\begin{align}\label{eqn:Paulicomm}
    &X_e (-1)^{f(a)}X_e=(-1)^{f(a+\tilde e)}~,\cr
    &Z_e (-1)^{h(b)}Z_e=(-1)^{h(b+\widetilde{e^\vee})}~,
\end{align}
where $\tilde s$ for $k$-cell $s$ is the indicator $k$-cochain that takes value $1$ on cell $s$ and 0 otherwise, and $f,h$ are arbitrary operator valued integer functions. 

Denote every edge adjacent to vertex $v$ by $e_i$ with $i$ labelling different edges, then $\sum_io(e_i)\tilde e_i=d\tilde v$ where $o(e_i)=\pm1$ is the relative orientation of edge $e_i$ and vertex $v$,\footnote{
Since $-1=1$ mod 2, it is irrelevant for the mod 2 relations discussed here and will be omitted below. They will be relevant in $\mathbb{Z}_N$ qudits, where the $\mathbb{Z}_N$ Gauss law operator is $G_v=\prod_{e:\partial e\supset v} X_e^{o(e)}$.}
since the latter is nonzero when evaluating on any edges that meet with the vertex $v$, $d\tilde v(01)=\tilde v(1)-\tilde v(0)$.
The first commutation relation in (\ref{eqn:Paulicomm}) implies
\begin{equation}\label{eqn:gaugeinvar}
    G_v (-1)^{f(a)}G_v=(-1)^{f(a+d\tilde v)}~.
\end{equation}
In other words, commuting with the local Gauss law operator ($X$-stabilizer) $G_v$ is equivalent to checking $f(a)$ is gauge invariant. This is the lattice statement of the field theory property that the Gauss law operator generates gauge transformations.

As an application of (\ref{eqn:gaugeinvar}), let us check the Wilson loop $W_\gamma$ commutes with the Gauss law operator $G_v$:
\begin{equation}
    G_v (-1)^{\oint_\gamma a} G_v=(-1)^{\oint_\gamma (a+d\tilde v)}=(-1)^{\oint _\gamma a}~.
\end{equation}
While the above commutation relation can also be evaluated by expressing the Wilson line in terms of $\prod Z_e$, the relations soon become complicated for more general logical gates, and the gauge-invariance is much easier and more systematic to check than gate identities. We will demonstrate further examples in Sec.~\ref{sec:commutationCnZ}.

\subsection{Copies of quantum codes on simplicial or cubical complexes and logical $C^{n-1}Z$ gates}
\label{sec:commutationCnZ}

\begin{figure}
    \centering
    \includegraphics[width=1\textwidth]{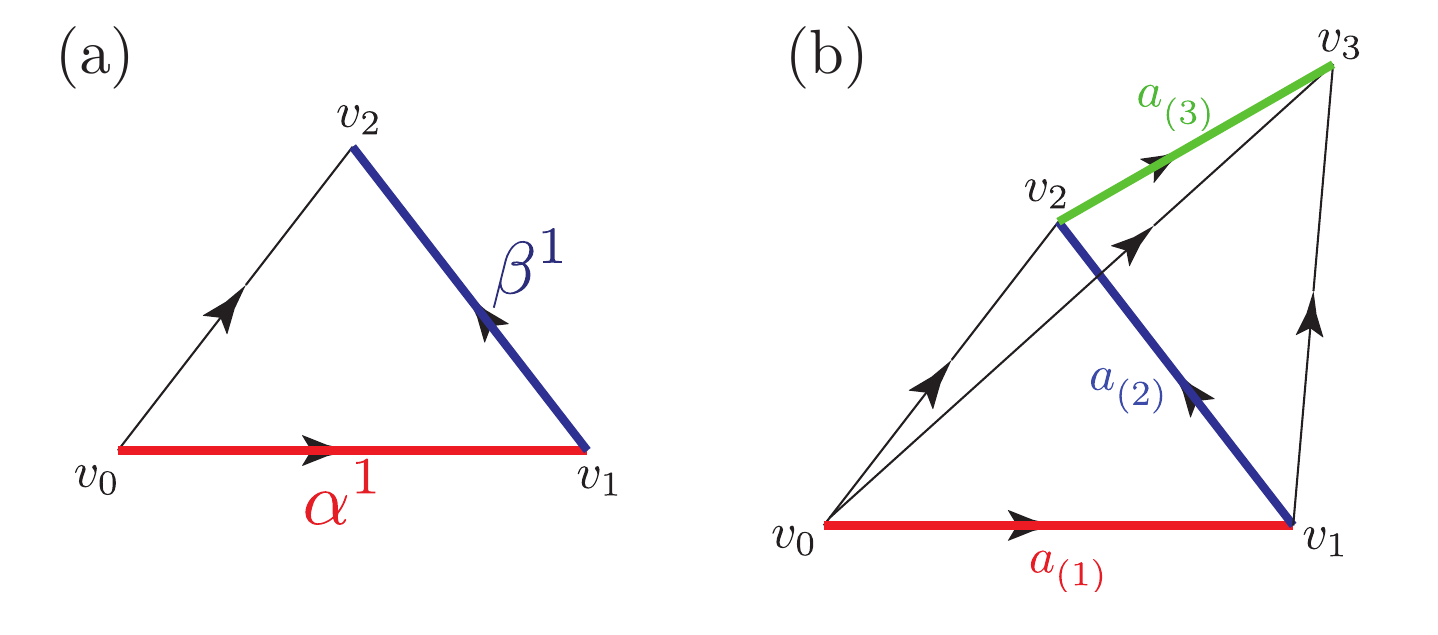}
    \caption{Definition of cup product on simplicial complexes.  The arrow points from vertices with lower order to vertices with higher order. (a)  Cup product of 1-cochains $\alpha^1\cup\beta^1$.
    For the 2D complex, one take product of cocycle values on the red and blue edges respectively in each 2-simplex. (b) Cup product of 1-cochains $a_{(1)}\cup a_{(2)}\cup a_{(3)}$. For the 3D complex, one takes the product of cocycle values on the red, blue, and green edges respectively.}
    \label{fig:cup_product_triangulation}
\end{figure}

\begin{figure}
    \centering
    \includegraphics[width=1\textwidth]{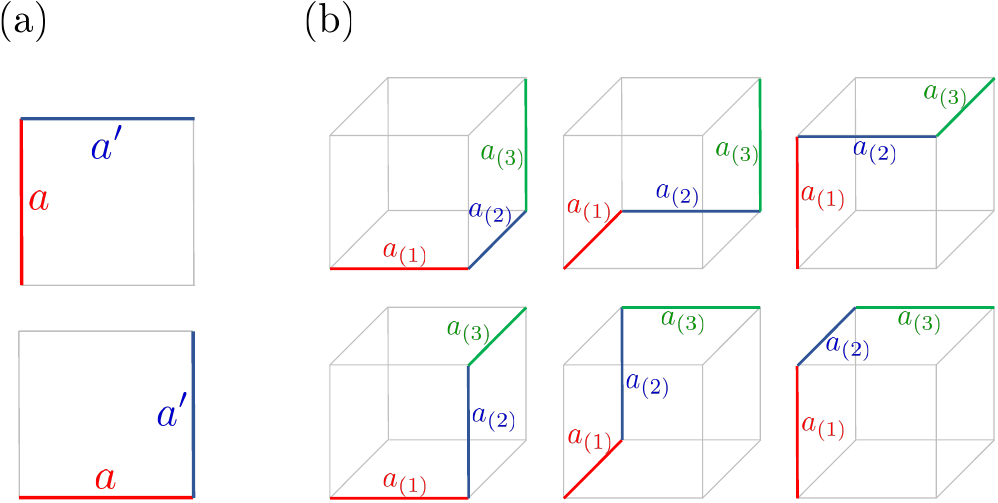}
    \caption{Definition of cup product on cubical complexes.  (a) In a 2D square complex, there are two paths contributing to the cup product. (b) In a 3D cubical complexes, there are six paths contributing to the cup product.
    }
    \label{fig:cup_product_cube}
\end{figure}

Let us use the formalism to show the cohomology operation via cup product  commutes with the $X$-stabilizers in the codes space (equivalent to Gauss law operator in the zero flux sector) and hence implements logical $C^{n-1}Z$  gate-- this provides a simple unifying way to reproduce and generalize known gate identities.

We first introduce cup products on a simplicial complex $\mathcal{L}$.  The vertices on the simplicial complex are assigned with a fixed global ordering.  
One can define the cup product `$\cup$' between a $p$-cochain $\alpha^p$ and $q$-cochain $\beta^q$ evaluated on a $(p+q)$-simplex $[v_0v_1\cdots v_{p+q}]$ as \cite{Hatcher:2001ut} 
\begin{align}\label{eq:cup_def}
 & (\alpha^p \cup \beta^q)([v_0v_1\cdots v_{p+q}]) \cr
=&\alpha^p([v_0v_1\cdots v_{p}])\beta^q([v_p v_{p+1}\cdots v_{p+q}])~, 
\end{align}
where the arguments contain the labels of ordered vertices $v_i$ with the ordering $v_0 \prec v_1 \prec v_2, \cdots \prec v_{p+q}$. An illustration of $p=q=1$ is illustrated in Fig.~\ref{fig:cup_product_triangulation}(a).
  
Now we introduce the unitary $U$ via cup products of the operator-valued 1-cocycles $a$ and $a'$ (gauge fields) on two identical copies of quantum codes defined on the simplicial complex $\mathcal{L}$:
\begin{equation}\label{eq:logical_CZ}
U = (-1)^{\oint a \cup a'}~,
\end{equation}
where $\oint$ represents a discrete sum over the 2-simplices in $\mathcal{L}$. In the specific case of homological codes defined on a triangulated manifold, this sum is over the triangulation $\mathcal{L}$. Using the definition in Eq.~\eqref{eq:cup_def}, we can hence re-express the unitary $U$ as
\begin{align}\label{eq:logical_gate_2D}
 U=& (-1)^{\int_{[v_0v_1v_2]} a([v_0,v_1])a'([v_1,v_2])} \cr
=&\prod_{[v_0 v_1 v_2]} \text{CZ}([v_0v_1],[v_1v_2])~,
\end{align}
which shows that $U$ is a constant-depth circuit composed of many CZ gates between qubits supported on the edges $[v_0v_1]$ and $[v_1v_2]$ in each 2-simplex.

We can also generalize the above cup product operation to a square lattice.
The corresponding logical gate is given by
\begin{equation}
   U= (-1)^{\oint a\cup a'}=\prod \text{CZ}^{(1,2)}_{\rightarrow,\uparrow}\text{CZ}^{(1,2)}_{\uparrow,\rightarrow}~,
\end{equation}
where the product is over all faces on the suqare complex, see Appendix \ref{sec:cupproductrev} for a review.\footnote{
For more details, see \cite{hatcher2002algebraic} for an introduction to the cup product $\cup$, and Appendix C of \cite{Chen:2021xuc} as well as \cite{Chen:2021ppt} for the definition of cup product on hypercubic lattice.
} 
On each face with corners given by $(x,y)$ coordinates $(0,0),(0,1),(1,0),(1,1)$,
$\text{CZ}^{(1,2)}_{\rightarrow,\uparrow}$ is the physical CZ gate acting on qubit 1 on the rightward edge $(0,0)\rightarrow (1,0)$ and qubit 2 on the upward edge $(1,0)\rightarrow (1,1)$. Similarly, $\text{CZ}^{(1,2)}_{\uparrow,\rightarrow}$ is the physical CZ gate acting on qubit 1 on the upward edge $(0,0)\rightarrow (0,1)$ and qubit 2 on the rightward edge $(0,1)\rightarrow (1,1)$. This rule is illustrated in Fig.~\ref{fig:cup_product_cube}(a).  In the specific case of a torus, the operator gives product of CZ logical gates for the two logical qubits on orthogonal 1-cycles. We note that the operator can be defined in homological code of any dimensions $d\geq 2$.

Let us check the operator commutes with the homological code Hamiltonians. The operator is diagonal in the Pauli $Z$ basis, and thus we only need to check whether it commutes with the Gauss law operators ($X$-stabilizers) $G_v,G_v'$ for the two copies of the homological code.
The commutator with the Gauss law operator $G_v$ is
\begin{equation}
    G_v(-1)^{\oint a\cup a'}G_v=(-1)^{\oint (a+d\tilde v)\cup a'}=(-1)^{\oint a\cup a'}(-1)^{\oint \tilde v da'}~.
\end{equation}
Thus it commutes with the Hamiltonian in the zero flux sector $(-1)^{da'}=1$. Similarly, one can show the operator commutes with the Gauss law operator $G_v'$ in the zero flux sector $(-1)^{da}=1$.

We now generalize the above cohomology operation to a triple cup product on 3D simplicial or cubical complexes in order to implement logical non-Clifford gates.  The unitary can be represented as
\begin{align}\label{eq:logical_CCZ}
U = (-1)^{\oint a_{(1)} \cup a_{(2)} \cup a_{(3)}}~,
\end{align}
where $a_{i}$ represents the 1-cocycle gauge field in the $i^\text{th}$ copy of codes. For simplicial 3-complex, the cup-product on each 3-simplex $[v_0,v_1, v_2, v_3]$  can be evaluated as
\begin{align}
  & (a_{(1)} \cup a_{(2)} \cup a_{(3)}) ([v_0,v_1,v_2, v_3]) \cr
  =& a_{(1)}([v_0,v_1])a_{(2)}([v_1,v_2])a_{(3)}([v_2,v_3])~,  
\end{align}
as illustrated in Fig.~\ref{fig:cup}(b).
The unitary $U$ can be re-expressed as
\begin{align}\label{eq:logical_gate_3D}
 U=& (-1)^{\int_{[v_0,v_1,v_2,v_3]} a_{(1)}([v_0,v_1])a_{(2)}([v_1,v_2])a_{(3)}([v_2,v_3])} \cr
=&\prod_{[v_0, v_1, v_2, v_3]} \text{CCZ}([v_0,v_1],[v_1,v_2],[v_2,v_3])~,
\end{align}
which explicitly shows the constant-depth circuit containing physical CCZ gates coupling the qubits in three copies of quantum codes.  The corresponding expressions for the case of cubic lattice is illustrated in Fig.~\ref{fig:cube}(b), where 6 paths contributes to the cup product in each cube.

The discussion can be generalized to $n$ copies of quantum codes with gauge fields $a_{(i)}$ for $i=1,\cdots,n+1$, and the logical operator:
\begin{equation}
 U=   (-1)^{\oint a_{(1)}\cup a_{(2)}\cup \cdots a_{(n)}}~.
\end{equation}
One can verify the operator commutes with the Hamiltonian in a similar way.

Now we derive the corresponding logical gate on a closed $(n+1)$-manifold $\M^{n+1}$.  In the code space, all the gauge fields are cocycles satisfying $da_{(i)}=0$. Therefore, we can re-express them using the 1-cocycle basis $\{\alpha^1_{(i)}\}$ of the cohomology group $H^1(M,\Z_2)$ for $i^\text{th}$ copy of codes: 
\begin{equation}
a_{(i)}=\sum_{\alpha^1_{(i)}}\hat{m}_{\alpha, i}\alpha^1_{(i)}~,
\end{equation}
where the quantum variable $\hat{m}_{\alpha, i}$ with eigenvalues $\{0,1\}$ are the winding numbers.  
We can hence re-express $U$ as
\begin{align}
 U &=  (-1)^{\sum_{\{\alpha^1_{(i)}\}} \oint (\hat{m}_{\alpha,1} \alpha^1_{(1)} )\cup (\hat{m}_{\alpha,2}  \alpha^1_{(2)}) \cup \cdots \cup (\hat{m}_{\alpha, n} \alpha^1_{(n)})}\cr
 \nonumber &  =\prod_{\{\alpha^1_{(i)}\}}\left[(-1)^{ \hat{m}_{\alpha,1} \hat{m}_{\alpha,2} \cdots \hat{m}_{\alpha, n}}\right]^{\oint {\alpha^1_{(1)} \cup \alpha^1_{(2)} \cup \cdots \cup \alpha^1_{(n)}}}\\
\nonumber   &=\prod_{\{\alpha^1_{(i)}\}} \overline{C^{n-1}Z}[\alpha^1_{(1)}, \alpha^1_{(2)}, \cdots, \alpha^1_{(n)} ]^{\oint {\alpha^1_{(1)} \cup \alpha^1_{(2)} \cup \cdots \cup \alpha^1_{(n)}}}~, \\
\end{align}
where the third equality has use the relation $(-1)^{ \hat{m}_1 \hat{m}_2 \cdots \hat{m}_{n}}$$=$$\overline{C^{n-1}Z}$.  Here, $\alpha^1_{(i)}$ serve as the labels of the logical qubits using the cocycle basis and the copy number $i$.  The logical $C^{n-1}Z$ gate between logical qubits is only non-trivial (not logical identity) if and only if $\oint {\alpha^1_{(1)} \cup \alpha^1_{(2)} \cup \cdots \cup \alpha^1_{(n)}}$ evaluates non-trivially.

\subsection{Addressable and parallelizable logical gates in qLDPC codes with higher-form symmetries}
In the above subsection, we focus on 0-form symmetries.  The disadvantage is the resulting logical $C^{n-1}Z$ gate applied to all the logical qubits and are hence not addressable in the context of qLDPC codes.   In Ref.~\cite{zhu2023non}, a strategy was proposed by using the $k$-form symmetries ($k>1$) to implement addressable logical gates.  It was shown that the following form can implement addressable logical CZ gates in two copies of 3D homological codes:
\begin{align}\label{eq:k-form_gate}
\nonumber   U(\Sigma_2) &= (-1)^{\oint_{\Sigma_2} a \cup a' } \equiv (-1)^{\oint a \cup a' \cup \Sigma^1 } \\
& = \prod_{\alpha^1,\beta^1}\overline{\text{CZ}}[\alpha^1, \beta^1]^{\oint \alpha^1 \cup \beta^1 \cup \Sigma^1}~,   
\end{align}
where $\Sigma_2$ is the 2-cycle (membrane) where the unitary is supported, and $\alpha^1$ is the Poincar\'e-dual 1-cocycle.  In addition, $\alpha^1$ and $\beta^1$ are the 1st cohomology bases of the two codes. As has been pointed out in Ref.~\cite{zhu2023non}, the number of independently addressable logical CZ gates (generators) correspond to the second Betti number 
\begin{equation}
b_2=\text{Rank}(H_2(\M^2; \ZZ_2))=b_1=k~, 
\end{equation}
which also equals to the first Betti number $b_1$ and hence the number logical qubits $k$ due to Poincar\'e duality.

Here, we also generalize the above result to addressable logical CCZ gate in three copies of 4D homological codes:
\begin{align}
U(\Sigma_3) &= (-1)^{\oint_{\Sigma_3} a_{(1)} \cup a_{(2)} \cup a_{(3)}}  \equiv (-1)^{\oint a_{(1)} \cup a_{(2)} \cup a_{(3)} \cup \Sigma^1}\cr & =\prod_{\alpha^1,\beta^1,\gamma^1}
\overline{\text{CCZ}}[\alpha^1, \beta^1, \gamma^1]^{\oint \alpha^1 \cup \beta^1 \cup \gamma^1 \cup \Sigma^1}~.
\end{align}
The number of independently addressable logical CCZ gate generators correspond to the third Betti number 
\begin{equation}
 b_3=\text{Rank}(H_2(\M^3; \ZZ_2))=b_1=k,    
\end{equation}
which again equals the first Betti number $b_1$ hence the number logical qubits $k$ due to Poincar\'e duality.

This can be further generalized to $q$-form symmetry in $n-q$ copies $n$D homological codes, which is supported on a codimension-$q$ ($n-q$ dimension) sub-complex or sub-manifold represented by a $q$-cycle $\Sigma_{n-q}$.  The symmetry can be expressed as
\begin{small}
\begin{align}
\nonumber & U(\Sigma_{n-q}) =  (-1)^{\oint_{\Sigma_{n-q}} a_{(1)}\cup a_{(2)}\cup \cdots a_{(n-q)}} \\  
&\equiv (-1)^{\oint  a_{(1)}\cup a_{(2)}\cup \cdots a_{(n-q)} \cup \Sigma^{k}} \cr
& = \prod_{\{\alpha^1_{(j)}\}}\overline{C^{n-q-1}Z}[\alpha^1_{(1)}, \alpha^1_{(2)}, \cdots, \alpha^1_{(n-q)} ]^{\oint {\alpha^1_{(1)} \cup \alpha^1_{(2)} \cup \cdots \cup \alpha^1_{(n-q)}} \cup \Sigma^q}~,\cr
\end{align}
\end{small}
where $\{\alpha^1_{(i)}\} \in H^1(M,\Z_2)$ represent the 1-cocycle bases for the   $i^\text{th}$ copy of codes which also serve as labels for the logical qubits and the $q$-cocycle $\Sigma^q$ is the Poincar\'e dual of the cycle $\Sigma_{n-q}$.   The number of independently addressable logical $C^{n-q-1}Z$ gate generators correspond to the $(n-q)^\text{th}$ or equivalently $q^\text{th}$ Betti number 
\begin{equation}
 b_{n-q}=\text{Rank}(H_{n-q}(M; \ZZ_2))=\text{Rank}(H_{q}(M; \ZZ_2))=b_q~.   
\end{equation}
Note that in the case that $q>1$, the number of logical generators $b_q$ no longer equals the logical qubit number $k=b_1$.
In general, the logical qubit and gate data is encoded in the chain complex as follows:
\begin{align}
&C_n \rightarrow \cdots \rightarrow C_{n-q} \rightarrow \cdots \rightarrow C_{2} \xrightarrow[]{\partial_{2}=\mathbf{H}_Z^T} C_1 \xrightarrow[]{\partial_{1}=\mathbf{H}_X} C_{0} \cr
&\overline{C^{n-1}Z}    \quad \ \overline{C^{n-q-1}Z}   \qquad \quad \  \overline{CZ} 
\qquad \quad \text{qubit} 
\end{align}
Hence, the logical qubit data  is encoded in the 1st chain group $C_1$ with the number being $k=b_1$, and the logical gate data in the highest Clifford hierarchy $\overline{C^{n-1}Z}$ is encoded in the chain group $C_n$. In the context of a homological code defined on a $n$-manifold $\M^n$ with a single connected component, we have the number of independent $\overline{C^{n-1}Z}$ being $b_n=b_0=1$ due to Poincar\'e duality and the fact that $b_0$ corresponds to the number of connected component in a manifold.  In order to get addressable and parallelizable logical non-Clifford gate, we have to use a $q$-form symmetry (with $q>0$), and the number of logical gate generator for $\overline{C^{n-q-1}Z}$ is determined by $b_q=b_{n-q}$. 

We note that although the above discussion focus on the case of homological codes, all the above formulae apply to the case of general simplicial or cell complexes as well except the Poincar\'e duality.

\subsection{$\mathbb{Z}_N$ higher-form homological codes: higher-form gauge theory}
\label{sec:ZNhigherform}
\subsubsection{Higher-form gauge theory}

In the previous subsections, we have focused on 1-form $\ZZ_2$ gauge theory, where the qubits are always placed on the edges, i.e., corresponding to the 1st chain group $C_1$. In this subsection, we consider the more general situation of a $q$-form $\ZZ_N$ gauge theory where the $N$-level qudits are placed on the $q$-cells and the $X$- and $Z$-stabilizers on the $(q-1)$-cell and $q$-cells respectively, as described by the following chain complex:
\begin{align}
&C_n \rightarrow \cdots \rightarrow C_{q+1} \xrightarrow[]{\partial_{q+1}=\mathbf{H}_Z^T} C_q \xrightarrow[]{\partial_{q}=\mathbf{H}_X} C_{q-1} \rightarrow \cdots \cr
&\qquad   \quad \quad \ Z\text{-stabilizer}  \qquad \   
 \text{qubit} \qquad X\text{-stabilizer}.
\end{align}
We call such a homological code an $n$-dimensional $q$-form homological code or a $(q,n-q)$ homological code where the two arguments specify the spatial dimensions of the logical-$Z$ ($q$-cycle) and -$X$ ($(n-q)$-cycle in the dual lattice) operators respectively.

The corresponding $q$-form gauge theory in $n$ spacetime dimension can be described by the action
\begin{equation}
    \frac{N}{2\pi}\int a_{q} db_{n-1-q}~,
\end{equation}
where $a_{q},b_{n-1-q}$ are $q$-form gauge field and $(n-1-q)$-form dual gauge field, and $(n-1)$ is the space dimension.
The theory has $\mathbb{Z}_N^{(n-1-q)}\times\mathbb{Z}_N^{(q)}$ higher-form symmetries generated by the Wilson operators $e^{iq_e\int a_{q}}$ and the magnetic operators $e^{iq_m\int b_{n-1-q}}$ for integers $q_e,q_m=0,1,\cdots N-1$, which have
Aharonov-Bohm braiding phase
\begin{equation}
    e^{2\pi i q_eq_m/N}~.
\end{equation}
We can parameterize the Hilbert space using the eigenbasis of either the Wilson operators or the magnetic operators, which correspond to Pauli $Z$ or $X$ basis respectively.

\subsubsection{Higher-form homological codes on a hypercubic lattice}

Let us consider $\mathbb{Z}_N$ homological code in $(n-1)$ space dimension, with $\mathbb{Z}_N$ qudit on $q$-simplices.
The higher-form homological code can be described by the Hamiltonian
\begin{equation}
    H=-\sum_{s_{q-1}} \prod_{\partial s_{q}\supset s_{q-1}} X^{o(s_{q})}_{s_{q}}-\sum_{s_{q+1}}\prod_{s_{q'}\subset \partial s_{q+1}} Z^{o(s_{q'})}_{s_{q'}}+\text{h.c.}~,
\end{equation}
where $o(s_{q})=\pm1$ depending on the orientation. 
The ground state subspace describes untwisted $q$-form $\mathbb{Z}_N$ gauge theory. The first term in the Hamiltonian imposes the $\mathbb{Z}_N$ Gauss law, while the second term ensures the $\mathbb{Z}_N$ $q$-form gauge field is flat.

On space $\Sigma_{n-1}$, the ground states have dimension $|H_{q}(\Sigma_{n-1},\mathbb{Z}_N)|=N^k$, where $k$ is the number of logical $\mathbb{Z}_N$ qudits.

The basic Wilson electric logical operator on closed $q$-dimensional submanifold $M_{q}$ is
\begin{equation}
    \overline{Z}(M_{q})=\prod_{s_{q}\subset M_{q}} Z_{s_{q}}~.
\end{equation}

The basic magnetic logical operator on closed $(n-1-q)$-dimensional submanifold $M_{n-1-q}$ on the dual lattice is
\begin{equation}
    \overline{X}(M_{n-1-q})=\prod_{s_{q}} X_{s_{q}}^{\#(s_{q},M_{n-1-q})}~,
\end{equation}
where $\#(s_{q},M_{n-1-q})$ is the (signed) intersection number.

\subsubsection{Code distance}
The smallest logical operator has size $L^{\text{min}(q,n-1-q)}$ where $L$ is the linear size of the system. The total number of physical qubits is $N\sim L^{n-1}$. Thus the code distance is
\begin{equation}
    d=\left\{
    \begin{array}{cl}
       O\left(N^{q/(n-1)}\right) & q\leq (n-1)/2 \\
         O\left(N^{(n-1-q)/(n-1)}\right)&  q\geq (n-1)/2
    \end{array}\right. ~,
\end{equation}
For $q=(n-1)/2$, the code distance is $d=O(N^{1/2})$.

\subsubsection{Commutation relations}

Let us specialize to the case of even $q$ or $N=2$.
Consider two copies of $q$-form homological codes with $q$-form gauge fields $a,a'$, the CZ logical operator on $2q$-dimensional cycle is:
\begin{equation}
    e^{{2\pi i\over N}\oint a\cup a'}~.
\end{equation}
In terms of physical gate, it is the product of control-Z operators.
Let us show that it commutes with the Hamiltonian. 
It is sufficient to show it commutes with the Gauss law operators on $(q-1)$-simplex $s_{q-1}$: for the Gauss law operator of the first qubit,
\begin{align}
    &G_{s_{q-1}}^\dag e^{{2\pi i\over N}\oint a\cup a'}
        G_{s_{q-1}}
    =e^{{2\pi i\over N}\oint a\cup a'}
    e^{{2\pi i\over N}\oint d\tilde s_{q-1}\cup a'}\cr
    &\qquad=e^{{2\pi i\over N}\oint a\cup a'}
    e^{{2\pi i\over N}\oint \tilde s_{q-1}\cup da'}~,
\end{align}
which equals $e^{{2\pi i\over N}\oint a\cup a'}$ in the zero flux sector $e^{{2\pi i\over N}da'}=1$. Similarly, the operator commutes with the Gauss law operator $G_{s_{q-1}}'$ for $a'$ in the zero flux sector $e^{{2\pi i\over N}da}=1$.

\subsection{Non-diagonal logical gates from cohomology operation}
\label{sec:nondiag1}
While the $C^nZ$ example above are diagonal logical gates, we can also construct non-diagonal gates.

Consider $\mathbb{Z}_2\times\mathbb{Z}_2$ 1-form homological code.
It has automorphism symmetry that maps the $\mathbb{Z}_2\times\mathbb{Z}_2$ generators $(U,U')$ to $(UU',U')$.
The automorphism is generated by the operator
\begin{equation}
    (-1)^{\int a'\cup b}~,
\end{equation}
where $a'$ is the second $\mathbb{Z}_2$ gauge field, and $b$ is the first dual $\mathbb{Z}_2$ gauge field. The dual gauge field lives on edges $e^\vee$ of the dual lattice, and the edge $e$ satisfy $e\cup e^\vee=1$ for $e,e^\vee$ intersects a point.\footnote{To be precise, the product $\cup$ between a cochain $a$ and a dual cochain $b$ is the chain-cochain pairing defined as $\int a\cup b := \int_{\mathrm{PD}(b)}a$. Here $\text{PD}(b)$ is a 1-chain Poincar\'e dual of $b$, defined as a sum of 1-chains intersecting with dual 1-chains on which $b$ is nonzero.}
Thus the operator on the lattice is
\begin{equation}
    (-1)^{\int a'\cup b}=\prod_e \text{CX}^{(2,1)}_{e}~,
\end{equation}
where $\text{CX}^{(2,1)}_{e}$ is the physical CNOT gate acting on the two qubits of the edge $e$, and the qubit 2 is the control qubit.
The operator gives the logical CNOT gate.

\section{Higher~Pontryagin~Powers:  Non-Clifford Logical Gates From SPT Beyond the Color Code Paradigm
}
\label{sec:higherPontryagin}

In this section, we will apply the formalism presented in Sec.~\ref{sec:reviewgaugefield} to obtain new non-Clifford fault-tolerant logical gates in $\mathbb{Z}_N$ higher-form homological code using the gauged SPT symmetry \cite{Barkeshli:2022edm}-- the symmetry operators are supported on submanifolds decorated with topological action of the $\mathbb{Z}_N$ higher-form gauge fields. Such topological actions are expressed in terms of cohomology operations of the $\Z_N$ gauge fields, and generalizes the logical gates reviewed in Sec.~\ref{sec:commutationCnZ} defined out of cup product.

We will focus on higher-form gauge fields of even degree $q=2r$, and the ``group cohomology SPTs" described by the group cohomology $H^n(B^{2r}\mathbb{Z}_N,U(1))$ for $2r$-form $\mathbb{Z}_N$ gauge field and SPT phases on $n$-dimensional submanifolds with $n\leq d$ (the space dimension). In table \ref{tab:SPT2-form}, we list the first few group cohomology $H^n(B^2G,U(1))$ for general finite Abelian group $G=\prod_i\mathbb{Z}_{N_i}$.

\begin{table}[t]
    \centering
    \begin{tabular}{|c|c|}
      \hline
        &  $H^m(B^2G,U(1))$ \\ \hline
     $m=2$ & $\prod_{i}\mathbb{Z}_{N_i}$\\
     $m=3$ & 1\\
      $m=4$   &  $\prod_i\mathbb{Z}_{\gcd(N_i,2)N_i}\times\prod_{i<j}\mathbb{Z}_{\gcd(N_i,N_j)}$ \\ 
      $m=5$ & $\prod_{i}\mathbb{Z}_{\gcd(N_i,2)}\times \prod_{i<j}\mathbb{Z}_{\gcd(N_i,N_j)}$ \\
      $m=6$ & $\prod_i\mathbb{Z}_{\gcd(3,N_i)N_i}\times \prod_{i\neq j}\mathbb{Z}_{\gcd(N_i,\gcd(2,N_j)N_j)}$
      \\ \hline
    \end{tabular}
    \caption{Classification of group cohomology SPT phases for 1-form symmetry, i.e. topological actions for general 2-form finite Abelian gauge group $G=\prod_i\mathbb{Z}_{N_i}$.}
    \label{tab:SPT2-form}
\end{table}

We will focus on higher-form homological code with even degree. The group cohomology SPT phases are not explicitly constructed except for the elements generated by Pontryagin square operations, i.e. the SPT phases with 1-form symmetry in 3+1d \cite{Gaiotto:2014kfa,Hsin:2018vcg,Tsui:2019ykk}. The logical gates for such SPT phases are Clifford gates as discussed in \cite{Chen:2021xuc,Hsin:2024pdi}. Here, we will generalize Pontryagin square to what we call Pontryagin powers, which generate the group cohomology $H^n(B^{2r}\mathbb{Z}_N,U(1))$. In particular, we will show that such cohomology operation produces new non-Clifford fault-tolerant logical gates for $\mathbb{Z}_N$ qudits.

 The ``higher Pontryagin power'' has been discussed in \cite{thomas1956generalization} as a generalization of Pontryagin square. We provide an explicit formula of the higher Pontryagin power using higher cup product \cite{Steenrod:1947} of $\Z_n$ cocycles, which is useful for physics application and leads to the new topological terms of higher-form gauge theories.

The discussion generalizes the logical gates from SPT phases in 1-form gauge theories \cite{Yoshida:2015boa,Barkeshli:2022wuz,Barkeshli:2022edm,Barkeshli:2023bta}. Unlike 1-form gauge theories, higher-form homological codes are self-correcting when the space dimension $d\ge 4$ and thus can be applied in single-shot quantum error correction \cite{Bombin:2014ksv} as demonstrated in recent experiment \cite{Berthusen:2024tsd}.

\subsection{Review of Pontryagin square SPT and logical gates}
\label{eqn:Pontryaginsquare}

SPT phases with $\mathbb{Z}_N$ 1-form symmetry in 3+1 spacetime dimensions are described by the 4d effective action
\begin{equation}
\text{Even }N :\quad    \frac{2\pi p}{2N}\int {\cal P}(B),\quad \text{Odd N}:\quad \frac{2\pi p'}{N}\int {\cal P}(B)~,
\end{equation}
where ${\cal P}$ is the Pontryagin square operation \cite{Whitehead1949}: for $\mathbb{Z}_N$ 2-form $B$ with even $N$ it produces a $\mathbb{Z}_{2N}$ valued 4-form, while for odd $N$ it produces a $\mathbb{Z}_N$ valued 4-form.
The SPT phases are used to construct new fault-tolerant logical CZ,S gates in 2-form homological code in \cite{Chen:2021xuc,Hsin:2024pdi}.

In this section, we will generalize the operation to higher power-- called Pontryagin power, as we construct in Appendix \ref{sec:higherP}.

\subsection{Logical gates from higher Pontryagin power SPT phases}
\label{sec:HPlogicalgates}

SPT phases with $\mathbb{Z}_N$ $(2r-1)$-form symmetry in $2nr$ spacetime dimension are generated by the effective action
\begin{equation}
    \frac{2\pi p}{Nn}\int {\cal P}(B;n)~,
\end{equation}
where $n$ is a divisor of $N$, and $B$ is a $\mathbb{Z}_N$ $2r$-form gauge field. The coefficient $p$ is an integer mod $Nn$.
The action has order $n$--if the gauge field is rescalled by $\ell$, the action will rescale by $\ell^n$. When $n=2$ and $r=1$, this is the Pontryagin square \cite{Whitehead1949}. 

In $\mathbb{Z}_N$ $2r$-form homological code in even space dimension $d$, we can construct logical gates using the higher Pontryain powers of $2r$-form gauge field $a$ or $(d-2r)$-form dual gauge field on lower-dimensional submanifolds in space. For logical gates that generalize the Pontryagin square, it is sufficient to focus on either the gauge field (as we will do here) or the dual gauge field-- the former requires $d>4r$ while the latter requires $4r>d$. When $d=4r$, there is also an electromagnetic duality, and in such case there is only SPT phases from Pontryagin square.

\subsubsection{Mixed terms}
We can also define mixed terms from the difference of Pontryain powers of the sum of the gauge field and the sum of the Pontryain powers of each individual gauge field. For example, 2-form gauge field for general finite Abelian gauge group $\prod_i \mathbb{Z}_{N_i}$ can have the topological term
\begin{equation}
    \frac{2\pi }{N\ell}\int {\cal P}\left(\sum_i\frac{N}{N_i}B^i;\ell\right)-\sum_i\frac{2\pi }{N\ell}\int {\cal P}\left(\frac{N}{N_i}B^i;\ell\right) ~,
\end{equation}
where $N$ is the least common multiple of $\{N_i\}$, and $\ell$ is a divisor of $N$.

For example, the $\mathbb{Z}_{N_1}$ 2-form gauge field $B$ and $\mathbb{Z}_{N_2}$ 2-form gauge field $B'$ can have the following mixed term: denote divisors $\ell_1|N_1$ and $\ell_2|N_2$, 
\begin{equation}\label{eqn:mixedtermP}
\frac{2\pi}{\gcd(\ell_1N_1,\ell_2N_2)}\int {\cal P}(B;\ell_1)\cup {\cal P}(B';\ell_2)~,
\end{equation}
which is a topological term of degree $2\ell_1+2\ell_2$. For example, when $\ell_1=1,\ell_2=2$ or $\ell_1=2,\ell_2=1$, the mixed topological terms in 5+1d are classified by $\mathbb{Z}_{\gcd(N_1,\gcd(2,N_2)N_2)}\times \mathbb{Z}_{\gcd(N_2,\gcd(2,N_1)N_1)}$.

Another example is when the power $\ell$ is a prime number and $N_i=N$. In such case, the mixed terms are given by ordinary cup product (see Appendix \ref{sec:refinecupproduct}).

\subsubsection{Higher Pontragin power operators using condensation presentation}
The logical gates from higher Pontryagin powers are mostly easily expressed in terms of $\mathbb{Z}_{N^2}$ homological code formalsim for $\mathbb{Z}_N$ homological code, where the latter is obtained by condensing $e^N$ (or $m^N$) with $e,m$ being the order-$N^2$ basic electric and magnetic excitations in $\mathbb{Z}_{N^2}$ homological code. We will focus on the  $e^N$ condensation in the following.
In Appendix \ref{sec:condesationZN} we review the formalism for condensation.
The idea is that starting with $\mathbb{Z}_{N^2}$ homological code, we measure the electric operators that create $e^N$. The resulting stabilizers are those that commute with these operators, and we also include the electric operators into the stabilizers. The resulting stabilizer Hamiltonian is
\begin{align}
    H'&=-\sum_{s_{2r-1}} \prod_{s_{2r-1}\subset \partial s_{2r}} X^N_{s_{2r}} -\sum_{s_{2r+1}} \prod_{s_{2r}\subset \partial s_{2r+1}} Z \cr &\qquad \qquad -\sum_{s_{2r}} Z^N_{s_{2r}}+\text{h.c.} ~.
\end{align}
The advantage of the formalism is that in terms of the parent $\mathbb{Z}_{N^2}$ $2r$-form gauge field, the higher Pontryagin powers reduce to simple cup product $a^n=a\cup a\cup \cdots \cup a$:
\begin{equation}
    U(M_{2rn})=e^{\frac{2\pi i}{Nn}\int {\cal P}(a;n)}=
    e^{\frac{2\pi i}{Nn}\int a^n}~.
\end{equation}
We will discuss examples of logical gates from higher Pontryagin powers in section  \ref{sec:HPlogicalgateexamples1} and \ref{sec:HPlogicalgateexamples2}.

\subsection{Family of new logical gates with increasing Clifford hierarchy}

The Pontryagin squares and the higher Pontryagin powers have the property that they are not nilpotent--we can iteratively apply them infinitely to construct new logical gates.

For example, the composition on $\mathbb{Z}_N$ 2-form gauge field $a$ with even $N$ gives
\begin{equation}
\frac{2\pi }{2^kN} \int    {\cal P}^k(a)~,
\end{equation}
where the iterative action maps $\mathbb{Z}_N$ 2-form to $\mathbb{Z}_{2N}$ 4-form, and to $\mathbb{Z}_{4N}$ 8-form, etc.
Similarly, the higher Pontragin powers can be applied iteratively.

The iterating operations give logical gates of increasing Clifford hierarchy. To see this, we note that the commutator of Pauli $X$ operator on the $k$-th iteration, which can be computed from twisted dimensional reduction on $S^{2}$ with $\oint a=1$, produces the operator with $k$ replaced by $(k-1)$:
\begin{equation}
    \frac{2\pi}{2^k N} \cdot 2^k\int{\cal P}^{k-1}(a)=\frac{2\pi}{2^{k-1}N}\int {\cal P}^{k-1}(a)~.
\end{equation}
For example, reducing the Pontryagin square action (\ref{eqn:Pontryaginsquare}) on $S^2$ with unit holonomy gives the Wilson operator $e^{{2\pi ip\over N}\int B}$, which is the logical Pauli $Z$ gate in 2-form homological code. Thus iteratively applying one additional higher Pontryagin power operation iteratively increases the Clifford hierarchy by one.

\subsection{New fault-tolerant logical $R_{k}$ gates}
\label{sec:HPlogicalgateexamples1}

Let us use the method to construct fault-tolerant logical $R_k$ gate,
\begin{equation}
    R_k=\mathrm{diag}(1,\exp({2\pi i}/{2^{k}}))~.
\end{equation}

Consider $\mathbb{Z}_N$ $2r$-form homological code, and the logical gate of SPT symmetry from the Pontryagin powers on space $M$ of various topology:
\begin{itemize}
\item $M=\mathbb{CP}^{n/2}$.  For $N=2,r=1$, when the space is $\mathbb{CP}^{n/2}$, $n$ is even.
There is a single logical qubit, since $H^2(\mathbb{CP}^{n/2},\Z_2)=\Z_2$. 

Here, $a$ is lifted into an integral cocycle $\tilde a\in Z^2(\mathbb{CP}^{n/2},\Z)$, which leads to $da=0$ mod $\Z$. 
The logical qubit can be labeled by the 2-form gauge field $a=m\omega$ where $\omega$ is the K\"ahler 2-form and $m=0,1$ labels the logical qubit.

We hence get $\mathcal{P}^k(a;2)=a^{2^k}$ mod $2^{k+1}$. 
For example, ${\cal P}(a;2)=a^2$ mod 4, ${\cal P}({\cal P}(a;2);2)=(a^2)^2=a^4$ mod 8, ${\cal P}^3(a;2)=(a^4)^2=a^8$ mod 16, and so on. Thus $\mathcal{P}^k(a;2)$ with $n\ge 2^{k+1}$ defines a logical $\overline{R_{k+1}}$ gate via the operator
\begin{equation}
    e^{{2\pi i\over 2^{k+1}}\int {\cal P}^k(a;2)}=e^{{2\pi i\over 2^{k+1}}[m]}=\overline{R}_{k+1}~,
\end{equation}
where the integral is over $\mathbb{CP}^{2^{k}}$ submanifold.

Note that the space dimension $n$ needs to grow exponentially to be at least $2^{k+1}$. This is not as efficient as the transversal $R_k$ gate in color code \cite{Kubica:2015mta}.

The code distance of 2-form homological code on $\mathbb{CP}^{n/2}$ can be derived from the bound on 2-systole on $\mathbb{CP}^{n/2}$ using Gromov's systolic inequalities \cite{Gromov:systole}. For $h$ a $2m$-cycle on $\mathbb{CP}^{n/2}$,  $\text{sup}_\omega \oint_h \omega \leq (m!/(n/2)!) \text{Vol}^{2m/n}$ \cite{Gromov:systole}, where Vol stands for the volume of the manifold. Thus for $m=1$, we find the area of 2-cycle on $\mathbb{CP}^{n/2}$ is bounded by $\text{Vol}^{2/n}/(n/2)!$. In other words, the code distance $d$ scales as the $2/n$ power of the total number of qubits (proportional to the volume).

\item $M=\mathbb{RP}^{n}$. For $N=2,r=1$, when the space is $\mathbb{RP}^n$.
There is a single logical qubit, since $H^2(\mathbb{RP}^{n},\Z_2)=\Z_2$. Denoting the first Stiefel-Whitney class $w_1$, generic element of second cohomology $a\in H^2(\mathbb{RP}^n,\Z_2)$ is given by $a=m w_1^2$ with $m=0,1$. The integral lift $\tilde w_1$ satisfies $d\tilde w_1=2\tilde{w}_1^2$, so $d(\tilde w_1^2)= 0$ mod $\Z$.  We hence get $\mathcal{P}^k(a,2)= a^{2^k}$ mod $\Z_{2^{k+1}}$. This again defines a 
logical $\overline{R_{k-1}}$ gate.

\item 
 For each divisor $m$ of $N$, consider the SPT operator with $m$th Pontryagin power on $M_{2rm}=\mathbb{C}\mathbb{P}^{rm}$. This gives ``$Z^{1/m}$'' logical gate, or $2\pi/(mN)$ rotation single $\mathbb{Z}_N$ qudit logical gate.

\end{itemize}

\subsection{New fault-tolerant logical $C^m R_k$ gate}
\label{sec:HPlogicalgateexamples2}

Similarly, let us use the method to construct fault-tolerant logical $C^mR_k$ gates, which are (multi-)control $R_k$ gates.

Using the mixed terms (\ref{eqn:mixedtermP}) from the higher Pontryagin powers we can construct logical $CR_k$ gates as follows. A simple example is provided by $N_1=N_2=2,r=1,\ell_1=\ell_2=2$ in 8 dimensional space of the topology $\mathbb{CP}^2\times \mathbb{CP}^2$, with 4 logical qubits labeled by the flat connection:
\begin{equation}
    B=n_1 \omega^1+n_2\omega^2,\quad B'=n'_1 \omega^1+n'_2\omega^2~,
\end{equation}
where $n_i=0,1,n_i'=0,1$ labels the 4 logical qubits, and
$\omega^1,\omega^2$ are the K\"ahler forms on the two $\mathbb{CP}^2$.
Integrating over the space picks up the $(\omega^1)^2(\omega^2)^2$ term, while all other terms integrates to zero.
Thus the SPT operator from the mixed term (\ref{eqn:mixedtermP}) is
\begin{equation}
    i^{\int {\cal P}(B)\cup {\cal P}(B')}=i^{[n_1n_2']}i^{[n_2n_1']}=\overline{CS}_{1,2'}\overline{CS}_{2,1'}~,
\end{equation}
where $[n]$ denotes $n$ mod 2. This gives a fault-tolerant logical CS gate.

More generally, consider $N_1=N_2=N$, $\ell_1=\ell_2=\ell> 1$ and $r=1$, and the space has topology $\mathbb{CP}^{\ell}\times\mathbb{CP}^{\ell}$, similar computation gives
\begin{align}
  &  e^{{2\pi i\over \ell N}\int {\cal P}(B;\ell)\cup {\cal P}(B';\ell)}\cr &=\exp\left( {2\pi i\over \ell N}\sum_{m,m'=0}^{\ell} n_1^m n_2^{\ell-m}n_1'^{\ell-m'}n_2'^{m'}
    \left(\begin{array}{c}
         \ell\\m
    \end{array}
    \right)
    \left(\begin{array}{c}
         \ell\\m'
    \end{array}
    \right)
    \right.\cr &\qquad \qquad \qquad \left.\cdot\int (\omega^1)^{m+\ell-m'}(\omega^2)^{\ell-m+m'}\right)\cr
    &=\exp\left(\frac{2\pi i}{\ell N}\sum_{m=0}^\ell   \left(\begin{array}{c}
         \ell\\m
    \end{array}
    \right)^2(n_1n_2')^m(n_2n_1')^{\ell-m}\right)~,
\end{align}
where $n_i,n_i'=0,1,\cdots,N-1$ label the logical $\mathbb{Z}_N$ qudits. For general $N,\ell$ this gives multi-control $R$ logical gates.

Let us illustrate this with several examples:
\begin{itemize}
    \item The case $N=2,\ell=2$ gives the logical gate
\begin{align}
&i^{[n_2][n_1']+[n_1][n_2']}(-1)^{n_1n_2'n_2n_1'}\cr 
&=(\overline{CR_2})_{n_2,n_1'}(\overline{CR_2})_{n_1,n_2'}(\overline{C^3Z})_{n_1,n_2',n_2,n_1'}~.
\end{align}

\item The case $N=4,\ell=4$ gives the logical gate
\begin{equation}
    \exp\; \frac{2\pi i}{16}\left([n_2][n_1']+[n_1][n_2']\right)\;i^{n_1n_2'n_2n_1'}~.
\end{equation}
\end{itemize}

We can also compose Pontryagin square iteratively. For example, let us consider $N=2$ on $(\mathbb{CP}^4)^4$, and
\begin{equation}\label{eqn:PP4}
    \exp\; \frac{2\pi i}{16}\int {\cal P}({\cal P}(B);4)\cup {\cal P}({\cal P}(B');4)~.
\end{equation}
We substitute 
\begin{equation}
    B=\sum_i n_i\omega^i,\quad B'=\sum_i n_i'\omega^i~,
\end{equation}
where $\omega^i$ for $i=1,2,3,4$ are the K\"ahler forms on the four copies of $\mathbb{CP}^4$, and $n_i,n_i'=0,1$ label the logical qubits. The integral collects the $(\omega^1)^4(\omega^2)^4(\omega^3)^4(\omega^4)^4$ term. Using Mathematica, we find the final answer
\begin{align}
    &i^{[n_1'][n_2'][n_3][n_4]}
    i^{[n_1'][n_2][n_3'][n_4]}
    i^{[n_1'][n_2][n_3][n_4']}\cr
    &\cdot i^{[n_1][n_2][n_3'][n_4']}
    i^{[n_1][n_2'][n_3'][n_4]+[n_1][n_2'][n_3][n_4']}\cr 
    &=(\overline{C^3R_2})_{n_1',n_2',n_3,n_4}
    (\overline{C^3R_2})_{n_1',n_2,n_3',n_4}\cr
    &\quad \cdot (\overline{C^3R_2})_{n_1',n_2,n_3,n_4'}
    (\overline{C^3R_2})_{n_1,n_2,n_3',n_4'}\cr
    &\quad \cdot (\overline{C^3R_2})_{n_1,n_2',n_3',n_4}
    (\overline{C^3R_2})_{n_1,n_2',n_3,n_4'}~.
\end{align}
Similarly, if instead of $(\mathbb{CP}^4)^4$ we consider $\mathbb{CP}^{16}$, there are two logical qubits labeled by the flat connections $B=n\omega$ and $B'=n'\omega$ where $\omega$ is the K\"ahler form of $\mathbb{CP}^{16}$ and $n,n'=0,1$. The topological response (\ref{eqn:PP4}) gives the logical gate
\begin{equation}
    e^{\frac{2\pi i}{16}[n][n']}=(\overline{CR_4})_{n,n'}~.
\end{equation}

The discussion can be generalized to other mixed topological responses, such as
\begin{equation}
    \exp\; \frac{2\pi i}{8}\int {\cal P}^2(B)\cup {\cal P}^2(B')~,
\end{equation}
where instead of the 4$^\text{th}$ Pontryagin power in (\ref{eqn:PP4}) we use only the Pontryagin squares. Then on $\mathbb{CP}^8$ with two logical qubits labeled by the flat connections $B=n\omega$ and $B'=n'\omega$ with $n,n'=0,1$ and $\omega$ being the K\"ahler form of $\mathbb{CP}^8$, this gives the logical gate
\begin{equation}
    e^{\frac{2\pi i}{8}[n][n']}=(\overline{CT})_{n,n'}~.
\end{equation}

\subsection{Non-diagonal logical gate}
\label{sec:nondiag2}
We can also construct non-diagonal logical gate, generalizing the discussion in Sec.~\ref{sec:nondiag1} using cohomology operation that involves both the gauge field and dual gauge field (chosen such that they can still be simultaneously diagonalized). The latter are gauge fields on the dual lattice.

As an example, consider 6-dimensional hypercubic lattice with $\mathbb{Z}_4$ 2-form and $\mathbb{Z}_2^2$ 4-form homological code.
Let us consider the logical gate corresponds to the unitary operator
\begin{equation}\label{eqn:aPb}
    i^{\int a\cup {\cal P}(b')}~,
\end{equation}
where $a$ is the $\mathbb{Z}_4$ 2-form gauge field, $b'$ is the $\mathbb{Z}_2$ 2-form dual gauge field.
In terms of physical gates, we can describe it by first going to the dual lattice, where $a$ is supported on every 4-cell. 
The operator is the product of control unitary operator $CU$, where the control physical $\mathbb{Z}_4$ qudit is on the 4-cell, and the unitary $U$ is the operator
\begin{equation}
    i^{\int {\cal P}(b')}
\end{equation}
on the 4-cell, for example as described in \cite{Chen:2021xuc}.
Then we dualize the lattice to the original lattice. Since hypercubic lattice has one-to-one correspondence between every $n$-cell on the original lattice and a $(6-n)$ cell on the dual lattice, this gives the physical gates expression for the unitary operator (\ref{eqn:aPb}).

What is the logical gate correspond to the operator (\ref{eqn:aPb})? For example, if we impose the boundary conditions such that the space has topology $T^2\times\mathbb{CP}^2$, where a flat 2-connection can be parameterized by $n\omega+m\omega'$ with integers $n,m$ for K\"ahler form $\omega'$ of $\mathbb{CP}^2$ and volume form $\omega$ of $T^2$, the operator evaluates to be
\begin{align}
\begin{split}
    &i^{\int (n_1\omega+m_1\omega')([n_2]\omega+[m_2]\omega')^2}=i^{n_1[m_2]}(-1)^{m_1[n_2][m_2]}\\
&=\overline{H}_{m_2}^{-1}\overline{CS}_{n_1,m_2}\overline{H}_{m_2} \\
    &\quad \cdot \overline{H}_{n_2}^{-1}\overline{H}_{m_2}^{-1}\overline{H}_{m_1}^{-1}\overline{CCX}_{n_2,m_2,m_1}\overline{H}_{m_1}\overline{H}_{m_2}\overline{H}_{n_2}\\ 
&=\overline{H}_{m_2}^{-1}\overline{CS}_{n_1,m_2}\overline{H}_{n_2}^{-1}\overline{H}_{m_1}^{-1}\overline{CCX}_{n_2,m_2,m_1}\overline{H}_{m_1}\overline{H}_{m_2}\overline{H}_{n_2}~,
\end{split}
\end{align}
where $n_1,m_1=0,1,2,3$ label the logical qudits from the $\mathbb{Z}_4$ homological code, and $n_2,m_2=0,1$ label the logical qubits from the $\mathbb{Z}_2$ homological code.

\subsection{Algorithm-efficient logical gates}

\begin{figure}
    \centering
    \includegraphics[width=1\textwidth]{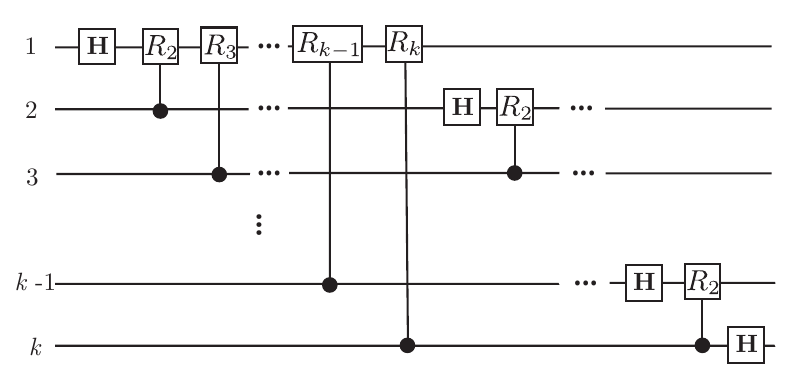}
    \caption{A logical circuit for the quantum Fourier transform algorithm which contains logical $CR_j$ gates ($2\le j \le k$).
    }
    \label{fig:QFT_circuit}
\end{figure}

A widely accepted way of performing universal fault-tolerant quantum computation is to construct a universal logical gate set and then compile the logical circuit using the gates in this gate set.  Widely used logical gate set includes the Clifford+$T$, CCZ+$H$, and so on.   Since the gate set is universal, an arbitrary unitary can be approximated efficiently with the given gates according to the Solovey-Kitaev theorem \cite{nielsen_chuang_2010}.  

On the other hand, for a specific class of quantum algorithms, it is possible that certain types of logical gates appear very frequently.  Therefore, a direct implementation of these gates fault-tolerantly would be desirable and can significantly optimize the logical circuit compiling.   We call such logical gates as \textit{algorithm-efficient fault-tolerant logical gates}.  

In Sec.~\ref{sec:HPlogicalgateexamples1} and \ref{sec:HPlogicalgateexamples2}, we have obtained the logical $R_k$ and $CR_k$ gates.  In many quantum algorithms, a small-angle rotation is needed.  Moreover, a lot of algorithms use controlled rotations, such as quantum Fourier transforms (QFT), which is key subroutine of Shor's algorithm, as well as quantum phase estimations and quantum simulation algorithms which needs a trotterized time evolution with the Hamiltonian as $e^{-i H \Delta t}$ \cite{nielsen_chuang_2010}.  For quantum machine learning, the quantum principle component analysis (QPCA) algorithm \cite{Lloyd:2014gcb} also needs such small-angle rotation. 

To be concrete, we consider a widely used QFT circuit as shown in Fig.~\ref{fig:QFT_circuit}.  There are $k$ logical qubits in this circuit,  and there needs controlled rotation from $CR_2$ to $CR_k$.   Using the type of $CR_k$ gates discussed in Sec.~\ref{sec:HPlogicalgateexamples2}, such logical circuit could be implemented more efficiently.   We leave the detailed study of such applications to future works.

\section{Non-Clifford Logical Gates from Boundary of SPT: Generalizing Folding}
\label{sec:logicalgatebdySPT}

Folding has been used to obtain logical gates such as logical $S$ gate from fold-transversal $CZ$ gate in the surface code, or the logical $T$ gate from the fold-transversal $T$ gate of the color code \cite{Kubica:2015mta}. Folding is generally interpreted as taking the copies of the codes, and then enforcing the gapped boundary conditions on the copies. The boundary is generally separated into a number of domains, each of which can realize distinct gapped boundary condition. Depending on the choices of the gapped boundaries, one can realize the distinct type of logical gates generated by the operators support at both bulk and boundary. 
We will develop a framework that illustrate the physics behind the procedure and generalize it to arbitrary gauge groups and new procedure to obtain new logical gates.

The framework obtains new logical gates from the boundary of logical gates whose expression in the bulk is given by SPT action in terms of cohomology operations. We will start by defining the boundary operation that can obtain boundary logical gates starting from parent bulk logical gate, and illustrate the relation with the folding procedure, and present examples of new non-Clifford logical gates obtained from our method.

\subsection{Definition of Boundary Operation ${\cal B}$}

Let us define a boundary operation ${\cal B}$ on group cochains with $U(1)$ coefficients as follows. The boundary operation is not nilpotent, and thus we can talk about the boundary of boundary--physically, this corresponds to higher-codimensional boundaries such as hinges or corners.

Starting from an SPT in $n$ dimension with $G$ symmetry, for instance described by group cocycle $\omega_n$ of $G$ with degree $n$ with $U(1)$ coefficients, this boundary operation ${\cal B}$ maps it to a $(n-1)$-cochain $\alpha_{n-1}(K,\omega_{n-1})$ in $(n-1)$ dimension specified by a subgroup $K$ and SPT in $(n-1)$ dimension $\omega_{n-1}$.
The subgroup $K$ satisfies the condition that $(\omega_n)|_K$ is an exact group cocycle, and $(\omega_n)|_K=d\alpha_{n-1}(K,\omega_{n-1})$, where different choices of $\alpha_{n-1}$ are related by $(n-1)$-dimensional SPT $\omega_{n-1}$ for the subgroup $K$: $\alpha_{n-1}'=\alpha_{n-1}+\omega_{n-1}$ with $d\omega_{n-1}=0$.
The boundary operation maps 
\begin{equation}
    {\cal B}^{(K,\omega_{n-1})}:\quad \omega_{n}\rightarrow \alpha_{n-1}(K,\omega_{n-1})~.
\end{equation}
Although we start with cocycle $\omega_n$, the boundary operation  can be extended to general cochains in $U(1)$ coefficients, and in particular it can be applied again on $\alpha_{n-1}$. Thus we can iteratively apply the boundary operation ${\cal B}$, and at each time it depends on a choice of subgroup and a group cocycle for the subgroup in one degree lower.

We remark that physically, the boundary operation uses the classification of gapped boundaries in finite group gauge theories, labeled by a subgroup that trivializes the bulk topological action, and a boundary topological action for the subgroup. See e.g. \cite{Beigi:2010htr,Hsin:2023ooo,Cordova:2024jlk} for examples.

The discussion applies to general finite group $G$ that can be Abelian or non-Abelian. For our purpose, we will focus on finite Abelian $G$. For reference, we list the classification of group cohomology SPT phases $H^n(BG,U(1))$ for the first few degrees in table \ref{tab:SPT1-form}.

\begin{table}[t]
    \centering
    \begin{tabular}{|c|c|}
 \hline
         & $H^m(BG,U(1))$ \\ \hline
      $m=1$   & $\prod_i\mathbb{Z}_{N_i}$ \\
      $m=2$  & $\prod_{i<j} \mathbb{Z}_{\gcd(N_i,N_j)}$  \\
      $m=3$ & $\prod_i \mathbb{Z}_{N_i}\times\prod_{i<j}\mathbb{Z}_{\gcd(N_i,N_j)}$ \\ 
       &\qquad  $\times \prod_{i<j<k}\mathbb{Z}_{\gcd(N_i,N_j,N_k)}$
      \\
      $m=4$ & 
      $\prod_{i<j}\mathbb{Z}_{\gcd(N_i,N_j)}^2\times\prod_{i<j<k}\mathbb{Z}_{\gcd(N_i,N_j,N_k)}^2$\\
      & \qquad $\times\prod_{i<j<k<l}\mathbb{Z}_{\gcd(N_i,N_j,N_k,N_l)}$ \\ 
      \hline
    \end{tabular}
    \caption{Classification of group cohomology SPT phases with ordinary unitary symmetry $G=\prod_i\mathbb{Z}_{N_i}$.}
    \label{tab:SPT1-form}
\end{table}

\subsubsection{Example of boundary operation ${\cal B}$: $G=\mathbb{Z}_2^2$}
\label{sec:exampleCZ->S}

To illustrate the boundary operation ${\cal B}$, let us consider the 2-cocycle $\omega_2=a\cup a'$ for $G=\mathbb{Z}_2\times \mathbb{Z}_2$ and $a,a'$ are the two independent $\mathbb{Z}_2$ 1-cocycles for the two $\mathbb{Z}_2$ gauge fields.

We will compute the image under the boundary operation with the diagonal subgroup $K=\mathbb{Z}_2$ generated by the diagonal element $(1,1)$ mod 2 of $\mathbb{Z}_2\times\mathbb{Z}_2=\{(m,n):m,n=0,1\}$, and with trivial 1-cocycle $\omega_1=0$ for $K$.
This subgroup sets $a=a'$ on the boundary, and the 2-cocycle $a\cup a'$ restricted to the subgroup becomes
\begin{equation}
    a\cup a'|_K=a\cup a=d [a]/2~,
\end{equation}
where $[a]$ is a lift of $a$ to $0,1$ in $\mathbb{Z}_4$ given by $a$ mod 2. Therefore $\alpha_1=[a]/2$. Thus we conclude that
\begin{equation}
    {\cal B}:\quad (-1)^{a\cup a'}\mapsto i^{[a]}~.
\end{equation}

\subsubsection{Example of boundary operation ${\cal B}$: $G=\mathbb{Z}_2^3$}
\label{sec:exampleCCZ->CS}

As another example, consider $G=\mathbb{Z}_2^3=\{(m,n,l):m,n,l=0,1\}$, and we start with the 3-cocycle $a\cup a''\cup a'$ where $a,a',a''$ are the three 1-cocycles for the three $\mathbb{Z}_2$ gauge fields.
We will compute the image under the boundary operation ${\cal B}$ for the subgroup $K=\mathbb{Z}_2^2=\{(m,n,m+n):m,n=0,1\}$ and trivial 2-cocycle for $K$. The subgroup sets $a''=a+a'$ mod 2. 
The 3-cocycle restricted to the subgroup becomes
\begin{align}
    &a\cup a''\cup a'|_K=a\cup (a+a')\cup a'=a^2\cup a'+a\cup a'^2 \cr &\quad =
    \frac{1}{2}\left(d[a]\cup [a']+[a]\cup d[a']\right)
    =\frac{d}{2}\left([a]\cup [a']\right)~,
\end{align}
where we used $[a]$ mod 2$=a$, $[a']$ mod 2$=a'$. Thus we conclude that
\begin{equation}
    {\cal B}:\quad (-1)^{a\cup a''\cup a'}\mapsto i^{[a]\cup [a']}~.
\end{equation}

We note that applying the boundary operation ${\cal B}$ again for the diagonal subgroup $K'=\mathbb{Z}_2=\{(m,m),m=0,1\}$ of $K=\mathbb{Z}_2^2$ and trivial 1-cocycle as in Sec.~\ref{sec:exampleCZ->S} gives
\begin{equation}
    {\cal B}^2:\quad (-1)^{a\cup a''\cup a'}\mapsto e^{\frac{2\pi i}{8}[a]}~.
\end{equation}

\subsubsection{Example: 2-form gauge field with $G=\mathbb{Z}_2^3$}
The discussion is not limited to ordinary gauge fields, but also apply to higher-form gauge fields. For example, take $G=\mathbb{Z}_2^3$ 2-form gauge fields $a,a',a''$ in 6 space dimensions, and the cocycle $a\cup a'\cup a''$.

We will consider the boundary map with the subgroup $\mathbb{Z}_2^2$ given by the condition $a+a'+a''=w_2$, where we note that we can modify the Dirichlet boundary condition with the second Stiefel-Whitney class $w_2$ (it means the string $\int (a+a'+a'')$ ending on the boundary gives rise to fermion particle on the boundary). To see this is a well-defined boundary condition, we note that the bulk cocycle is trivialized:
\begin{align}
    (-1)^{a\cup a'\cup (a+a'+w_2)}&=(-1)^{Sq^2(a\cup a')+w_2\cup a\cup a'+Sq^1a Sq^1 a'}\cr
    &=(-1)^{d(aSq^1 a')/2}~,
\end{align}
where we used $Sq^1 a=da/2$.
Thus the boundary operation gives 
\begin{equation}
    {\cal B}:\quad (-1)^{a\cup a'\cup a''}\mapsto i^{[a]\cup d[a']/2}~.
\end{equation}

\subsection{Logical gates from boundary operations}

The above boundary operation leads to the logical gates of the stabilizer code with gapped boundaries. Suppose that the bulk stabilizer code is given by the $G$ gauge theory in $n$ spatial dimensions, with a gapped boundary with the breaking of the gauge group $G\to K$. Then, the bulk-boundary system has a 0-form global symmetry generated by
\begin{align}
    \exp\left(2\pi i \int_{\text{bulk}}\omega_n - 2\pi i\int_{ \text{bdry}}\mathcal{B}^{K,\omega_{n-1}}(\omega_n) \right)~,
\end{align}
where $\omega_n\in Z^{n}(G,U(1)), \omega_{n-1}\in Z^{n-1}(K,U(1))$.
This becomes a logical gate of the stabilizer code with a gapped boundary.

Such a construction of the logical gate also applies when the boundary is separated into a number of domains and each domain realizes a distinct gapped boundary, e.g., a surface code. In that case the boundary domain further has its boundary, e.g., hinges or corners. The hinge or corner supports the topological action given by the iterated boundary operations.

\begin{figure*}[t]
    \centering
    \includegraphics[width=0.5\textwidth]{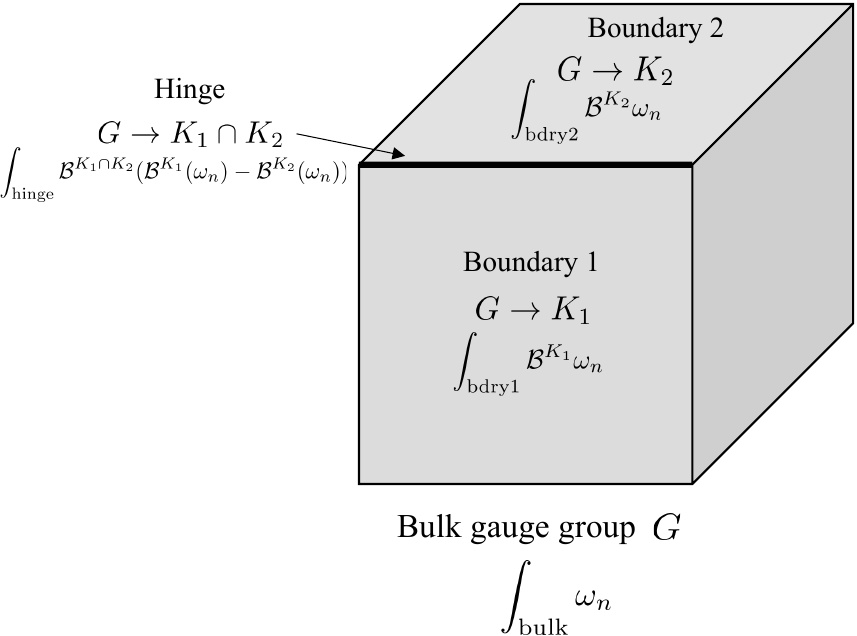}
    \caption{The logical gate of the $G$ gauge theory in the presence of gapped boundaries and hinges. 
    }
    \label{fig:hinge}
\end{figure*}

For instance, suppose that the boundary is separated into two regions 1 and 2 along a $(n-2)$-dimensional hinge, and each domain breaks the gauge group as $G\to K_1, G\to K_2$, and the hinge breaks the gauge group as $G\to K_1\cap K_2$. Then there is a 0-form symmetry generated by (see Figure \ref{fig:hinge})
\begin{align}
\begin{split}
    &\exp\left(2\pi i \int_{\text{bulk}}\omega_n - 2\pi i\int_{ \text{bdry}_1}\mathcal{B}^{K_1,\omega_{n-1}}(\omega_n)\right.\cr&\left.\qquad \qquad \qquad  +2\pi i\int_{ \text{bdry}_2}\mathcal{B}^{K_2,\omega'_{n-1}}(\omega_n)
     \right) \\
    \times & \exp\left(2\pi i \int_{\text{hinge}} \mathcal{B}^{K_1\cap K_2,\omega_{n-2}}(\mathcal{B}^{K_1,\omega_{n-1}}(\omega_n)) \right.\cr&\left.\qquad \qquad \qquad  
    - \mathcal{B}^{K_1\cap K_2,\omega'_{n-2}}(\mathcal{B}^{K_2,\omega'_{n-1}}(\omega_n)) \right)~.
    \end{split}
\end{align}

\subsection{Fault-tolerant logical $R_N$ gate in $N$ dimensional $\Z_2^N$ homological code: $N=2,3,4,5$}

\begin{figure*}[t]
    \centering
    \includegraphics[width=0.8\textwidth]{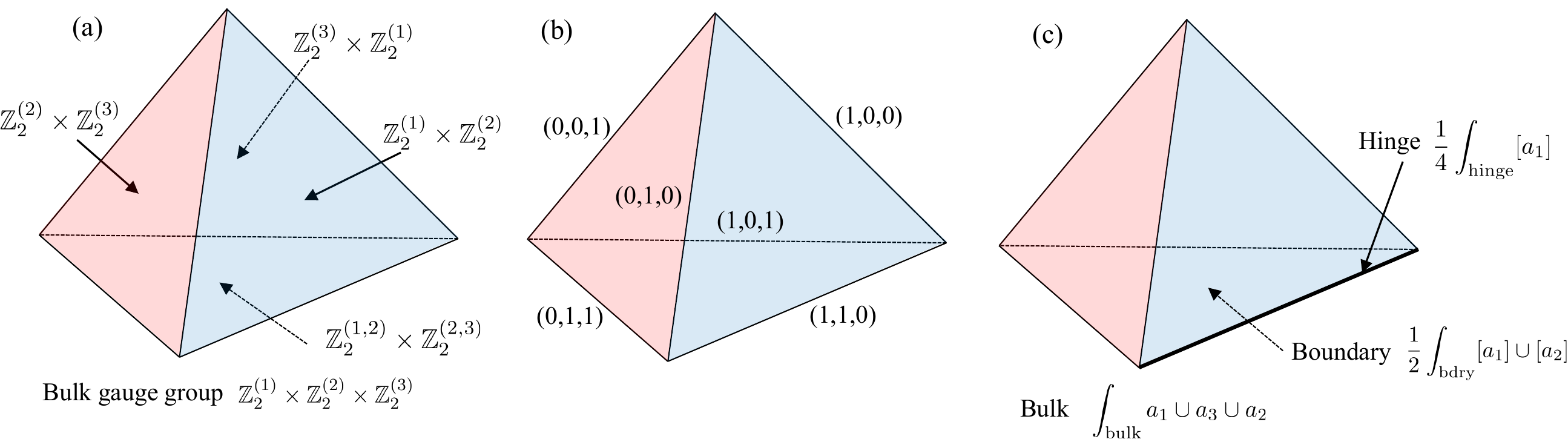}
    \caption{The gapped boundary conditions for the $\Z_2^3$ gauge theory on a tetrahedron ($N=3$). (a): Patterns of the broken gauge groups at each boundary face. (b): The holonomy of a single non-trivial $\Z_2^3$ gauge field compatible with the boundary conditions. This configuration of $\Z_2^3$ gauge fields correspond to the logical state $\ket{1}$. (c): The logical gate support at the bulk, a single boundary face at the bottom, and a single hinge edge of a tetrahedron. This defines a (conjugate of) the logical $T$ gate.
    }
    \label{fig:tetra}
\end{figure*}

The above construction from boundary operations leads to the $R_N=\mathrm{diag}(1,\exp({2\pi i}/{2^{N}}))$ logical gate of $\Z_2^N$ homological code in $N$ spatial dimensions, with $N=2,3,4,5$.
When $N=2$, it is equivalent to the $S$ gate in the folded surface code in 2D. When $N=3$, it is the $T$ gate of the $\Z_2^3$ homological code with gapped boundaries. See Figure \ref{fig:tetra} for the illustration of $N=3$. We give an explicit construction of the $R_N$ gate with $N=2,3,4,5$, and conjecture that our construction can be extended for the $R_N$ gate with generic $N$.

We note that the logical $R_N$ gate of the color code on a hyper tetrahedron has been obtained in \cite{Kubica2015universal, bombin2015gaugecolorcodesoptimal} by the transversal $R_N$, and the $C^{N-1}Z$ gate of the homological code on a hypercube has been obtained in \cite{Kubica:2015mta} by the transversal $C^{N-1}Z$ gate. Meanwhile the logical $R_N$ gate of the $\Z_2^N$ homological code with $N=3,4,5$ is new, implemented by a finite depth circuit but not transversal. 

Let us consider the $\Z_2^N$ homological code on a hyper tetrahedron, namely a $N$-simplex $(01\dots N)$. Its boundary domains are $(N-1)$-simplices, and hinges are $(N-2)$-simplices, and so on. 
The $(N-1)$-simplices realize the gapped boundary given by the breaking of the gauge group.
The breaking pattern of $\Z_2^N=\Z_2^{(1)}\times\dots\times \Z_2^{(N)}$ at the $(N-1)$-simplices are given as follows (see Figure \ref{fig:tetra} (a)):
\begin{itemize}
    \item On a $(N-1)$-simplex $(01\dots \hat{j}\dots N)$ with $1\le j\le N$, the gauge group breaks as $\Z_2^{N-1}=(\Z_2^{(1)}\times\dots\times \Z_2^{(N)})/\Z_2^{(j)}$.
    \item On a $(N-1)$-simplex $(12\dots N)$, the gauge group breaks as $\Z_2^{N-1}=\Z_2^{(1,2)}\times\Z_2^{(2,3)}\times\dots \times \Z_2^{(N-1,N)}$. Here $\Z_2^{(j,k)}:= \text{diag}(\Z_2^{(j)},\Z_2^{(k)})$.
    \end{itemize}
The breaking pattern at simplices with lower degrees are iteratively determined by requiring that $(j-1)$-simplices $\Delta_{j-1}$ breaks the gauge group as $K\cap K'\cap K''\dots $ if the $j$-simplices $\{\Delta_j, \Delta'_j, \Delta''_j\dots \}$ whose boundary contains $\Delta_{j-1}$ breaks the gauge group as $K, K', K''\dots $ respectively. Concretely, this leads to the breaking pattern as
\begin{itemize}
    \item On a $k$-simplex $(0 a_1\dots a_{k})$ with $0< a_1<\dots a_{k}$, the gauge group breaks as $\Z_2^{k}=\Z_2^{(a_1)}\times\dots\times \Z_2^{(a_k)}$.
    \item On a $k$-simplex $(a_{1}\dots a_{k+1})$ with $0< a_1<\dots a_{k+1}$, the gauge group breaks as $\Z_2^{k}=\Z_2^{(a_1,a_2)}\times\Z_2^{(a_2,a_3)}\times\dots \times \Z_2^{(a_{k},a_{k+1})}$. 
    \end{itemize}

\subsubsection{Boundary and hinge actions for phase gates}
We consider the 0-form symmetry in the bulk generated by the $N$-cocycle $\omega_N= a_1\cup\dots\cup  a_{N}+d\chi$ with a certain cochain $\chi$. On the boundary simplices with degree $k$, the action is given by the iterated boundary operations $\mathcal{B}^{N-k}(\omega_N)$. The actions on the simplices are explicitly given for $N=2,3,4,5$ in the following:

\vspace{3mm}
\noindent{\bf The actions for $N=2$: } The action for the logical $S$ gate with $N=2$ is given as follows:
\begin{itemize}
    \item On a 2-simplex $(012)$, there is a bulk action $\omega_2=a_1\cup a_2$.
    \item On a 1-simplex $(12)$, there is a boundary action $\mathcal{B}(\omega_2) = [a_1]/2$. The actions on other 1-simplices are zero.
\end{itemize}
The logical gate is then given by
\begin{align}
    U_2 = \exp\left(\pi i \int_{(012)}\omega_2 - \pi i\int_{(12)}\mathcal{B}(\omega_2)\right)~.
    \label{eq:gateonsimplex N=2}
\end{align}

\vspace{3mm}
\noindent{\bf The actions for $N=3$: } The action for the logical $T$ gate with $N=3$ is given as follows:
\begin{itemize}
    \item On a 3-simplex $(0123)$, there is a bulk action $\omega_3=a_1\cup a_3\cup a_2$.
    \item On a 2-simplex $(123)$, there is a boundary action $\mathcal{B}(\omega_3)= [a_1]\cup[a_2]/2$. The actions on other 2-simplices are zero.
    \item On a 1-simplex $(12)$, there is a hinge action $\mathcal{B}^2(\omega_3)= [a_1]/4$. The actions on other 1-simplices are zero.
\end{itemize}
The logical gate is then given by (see Figure \ref{fig:tetra} (c))
\begin{align}
    U_3 = \exp\left(\pi i \int_{(0123)}\omega_3 - \pi i\int_{(123)}\mathcal{B}(\omega_3) +\pi i\int_{(12)}\mathcal{B}^2(\omega_3)\right)~.
    \label{eq:gateonsimplex N=3}
\end{align}

\vspace{3mm}
\noindent{\bf The actions for $N=4$: } The action for the logical $R_4$ gate with $N=4$ is given as follows:
\begin{itemize}
    \item On a 4-simplex $(01234)$, there is a bulk action $\omega_4=a_1\cup a_4\cup a_3\cup a_2$. 
    \item On a 3-simplex $(1234)$, there is a boundary action $\mathcal{B}(\omega_4) = [a_1][a_3][a_2]/2 + a_1(a_2\cup_1 a_3) a_2$, where the product of $a_j$ denotes the cup product.
   The higher cup product $\cup_1$ is reviewed in appendix \ref{sec:higherP}. 
    This can be evaluated by plugging the boundary condition $a_4=a_1+a_2+a_3$ into the expression of $\omega_4$, then $\omega_4$ becomes a coboundary $\omega_4 = d(\mathcal{B}(\omega_4))$.
    The actions on other 3-simplices are zero.
    \item On a 2-simplex $(123)$, there is a hinge action $\mathcal{B}^2(\omega_4)$. Since $a_3=a_1+a_2$ mod 2 on the simplex $(123)$, the boundary map is explicitly evaluated by
    \begin{align}
    \begin{split}
        d\mathcal{B}^2(\omega_4) =& \mathcal{B}(\omega_4)|_{a_3=a_1+a_2} \\
        =& \frac{1}{2}[a_1]([a_1]+[a_2]+2a_2\cup_1a_1)[a_2] \\
        & + a_1(a_2\cup_1 (a_1+a_2)) a_2 \\
        =& \frac{1}{2}([a_1]^2[a_2]+[a_1][a_2]^2) + a_1a_2a_2 \\
        =& \frac{1}{2}([a_1]^2[a_2]-[a_1][a_2]^2) \\
        =& \frac{1}{4}d([a_1][a_2])~.
        \end{split}
    \end{align}
    Therefore $\mathcal{B}^2(\omega_4)= \frac{1}{4}[a_1]\cup[a_2]$ on a simplex (123). The actions on other 2-simplices are zero.
    \item On a 1-simplex $(12)$, there is a corner action $\mathcal{B}^2(\omega_3)= [a_1]/8$. The actions on other 1-simplices are zero. 
    
\end{itemize}
The logical gate is then given by
\begin{align}
    U_4 &= \exp\left(\pi i \int_{(01234)}\omega_4 - \pi i\int_{(1234)}\mathcal{B}(\omega_4) \right.\cr &\left.\quad -\pi i\int_{(123)}\mathcal{B}^2(\omega_4) + \pi i\int_{(12)}\mathcal{B}^3(\omega_4)\right)~.
    \label{eq:gateonsimplex N=4}
\end{align}
\vspace{3mm}

\noindent{\bf The actions for $N=5$: } The nonzero action for the logical $R_5$ with $N=5$ is given as follows:
\begin{itemize}
    \item On a 5-simplex $(012345)$, the action is $\omega_5=a_1\cup a_5\cup a_4\cup a_3\cup a_2$. 
    \item On a 4-simplex $(12345)$, the action is $\mathcal{B}(\omega_5) = [a_1][a_4][a_3][a_2]/2 + a_1(a_3\cup_1 a_4) a_3a_2 + a_1(a_2\cup_1 a_4)a_3a_2 + a_1a_4(a_2\cup_1a_3)a_2$.
    \item On a 3-simplex $(1234)$, the action is $\mathcal{B}^2(\omega_5) = [a_1][a_3][a_2]/4 - [a_1]([a_2]\cup_1[a_3])[a_2]/2$.
    \item On a 2-simplex (123), the action is $\mathcal{B}^3(\omega_5) = [a_1][a_2]/8$.
    \item On a 1-simplex (12), the action is $\mathcal{B}^4(\omega_5) = [a_1]/16$.
\end{itemize}
The action on the other simplices are zero. The logical gate is given by
\begin{small}
\begin{align}
    U_5 &= \exp\left(\pi i \int_{(012345)}\omega_5 - \pi i\int_{(12345)}\mathcal{B}(\omega_4) \right. \cr &\left.\; +\pi i\int_{(1234)}\mathcal{B}^2(\omega_4)+ \pi i\int_{(123)}\mathcal{B}^3(\omega_4) -\pi i\int_{(12)}\mathcal{B}^4(\omega_4)\right)~.
    \label{eq:gateonsimplex N=5} \cr
\end{align}
\end{small}

\subsubsection{Nontrivial $\Z_2^N$ gauge field on a simplex and a logical qubit}

Let us get back to the discussion for generic $N$.
The code space on this $N$-simplex is a single qubit. 
The states of the logical qubit $\ket{0}, \ket{1}$ correspond to the configurations of $\Z_2^N$ gauge field on the $N$-simplex compatible with the boundary conditions. The state $\ket{0}$ corresponds to the trivial $\Z_2^N$ gauge field.
The state $\ket{1}$ corresponds to the nontrivial $\Z_2^N$ gauge field, labeled by the holonomies along the 1-simplices $(jk)$ which is given as follows (see Figure \ref{fig:tetra} (b)):
\begin{itemize}
    \item On a 1-simplex $(0j)$, the holonomy of $\Z_2^{(j)}$ is nontrivial, and it becomes $(0,\dots, 1,\dots 0)\in \Z_2^N$ where $j$-th entry is 1 otherwise 0.
    \item On a 1-simplex $(jk)$ with $0<j<k$, the holonomy of $\Z_2^{(j,k)}$ is nontrivial, and it becomes $(0...1...1...0)\in \Z_2^N$ where the $j,k$-th entries are 1 otherwise 0.
\end{itemize}
One can check that the above $\Z_2^N$ gauge field is the flat $\Z_2^N$ gauge field on a $N$-simplex compatible with the symmetry breaking pattern, in the sense that the holonomy $k\in\Z_2^N$ is an element of the unbroken group $K$ on 1-simplices.

\subsubsection{The action of the logical gate}
The expression of the logical gate can be obtained by evaluating the sum of cocycles in \eqref{eq:gateonsimplex N=2}, \eqref{eq:gateonsimplex N=3}, \eqref{eq:gateonsimplex N=4}, \eqref{eq:gateonsimplex N=5} on the $2,3,4,5$-simplex with the gauge field that corresponds to $\ket{0},\ket{1}$ respectively. The action on $\ket{0}$ obviously becomes a trivial phase. The action on $\ket{1}$ becomes the following phases:
\begin{itemize}
    \item The action of $U_2$ on  $\ket{1}$ is the phase $\exp\left(\pi i - \frac{\pi i}{2}\right) = i$. 
    \item The action of $U_3$ on $\ket{1}$ is the phase $\exp\left(0 - \frac{\pi i}{2} + \frac{\pi i}{4}\right) = e^{-\frac{\pi i}{4}}$.
    \item The action of $U_4$ on $\ket{1}$ is the phase $\exp\left(0 -0 - \frac{\pi i}{4} +\frac{\pi i}{8} \right) = e^{-\frac{\pi i}{8}}$.
    \item The action of $U_5$ on $\ket{1}$ is the phase $\exp\left(0 -0 +0 + \frac{\pi i}{8} -\frac{\pi i}{16} \right) = e^{\frac{\pi i}{16}}$.
\end{itemize}
Therefore the logical gates implement $U_2=S, U_3= T^\dagger, U_4 = R_4^\dagger$, $U_5 = R_5$.

\subsubsection{Example: $T$ gate of the 3D $\Z_2^3$ homological code on a cube}

For instance, the above construction of the $R_3=T$ gate directly leads to the $T$ gate on the three copies of $\Z_2$ homological codes on a cube. This construction can be generalized to the $R_N$ gate on $N$ copies of the $\Z_2$ homological code on a $N$ dimensional hypercube with $N=4,5$.

\begin{figure*}[t]
    \centering
    \includegraphics[width=0.6\textwidth]{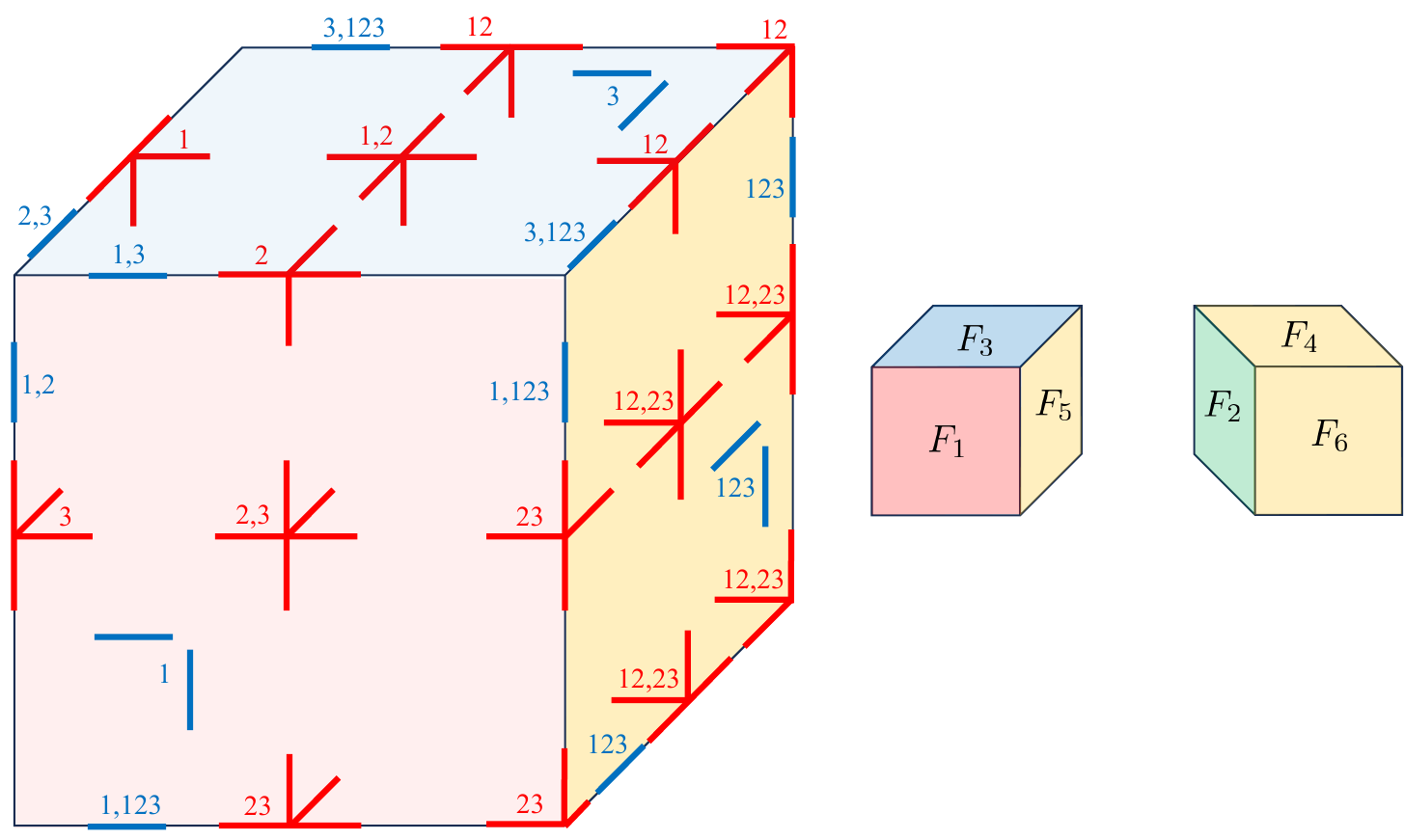}
    \caption{The boundary stabilizer of the three copies of $\Z_2$ homological codes on a cube. The four types of the gapped boundaries correspond to the colors Red, Green, Blue, Yellow of the faces. The labels of the faces are assigned according to the standard dice, i.e., the opposite faces have a pair of labels $F_j, F_{7-j}$. The $X$ stabilizers at the boundary or corner are defined by truncation of the bulk stabilizer. The label such as $jk$ at the stabilizer means we add the term $A_v^{(j)}A_v^{(k)}$ to the stabilizer group. The boundary or corner stabilizers hidden in the back of the figure (such as the ones on $F_2, E_{24}...$) can be determined by the cyclic permutation of the boundary labels R$\to$G$\to$B$\to$R associated with the labels of the stabilizers $1\to 2 \to 3\to 1$.
    }
    \label{fig:cube}
\end{figure*}

The cube has the six faces $F_1,\dots,F_6$ on its boundary. The labels 1,\dots, 6 of the faces are assigned according to that of the standard dice (see Figure \ref{fig:cube}). Following the above discussions, the boundary condition on each face can be assigned as follows:
\begin{itemize}
    \item On a face $F_1$, we condense the magnetic loops $m_2,m_3$ and an electric particle $e_1$. This corresponds to the unbroken gauge group $\Z_2^{(2)}\times \Z_2^{(3)}$ at this boundary.
    \item On a face $F_2$, we condense the magnetic loops $m_3,m_1$ and an electric particle $e_2$.
    \item On a face $F_3$, we condense the magnetic loops $m_1,m_2$ and an electric particle $e_3$.
    \item On faces $F_4,F_5,F_6$, we condense the magnetic loops $m_1m_2,m_2m_3$ and an electric particle $e_1e_2e_3$. This corresponds to the unbroken gauge group $\Z_2^{(1,2)}\times \Z_2^{(2,3)}$ at these boundaries.
\end{itemize}
The edge of the cube adjacent to the faces $F_j,F_k$ is denoted by $E_{jk}$. The condensed objects at $E_{jk}$ is given by the magnetic loops condensed at both $F_j$ and $F_k$ at the same time, and the electric particles that braid trivially with them. For instance, the edge $F_{14}$ condenses $m_2m_3,e_1,e_2e_3$. 

Three copies of $\Z_2$ homological codes on the cube then has the following Hamiltonian (see Figure \ref{fig:cube})
\begin{small}
\begin{align}
\begin{split}
    &H =  -\sum_{v\in \text{cube}/F} (A_v^{(1)} + A_v^{(2)} + A_v^{(3)}) \cr &- \sum_{p\in \text{cube}/F} (B_p^{(1)} + B_p^{(2)} + B_p^{(3)}) \\
    & -\sum_{v\in F_1/E} (A_v^{(2)} + A_v^{(3)}) - \sum_{v\in F_2/E} (A_v^{(3)} + A_v^{(1)}) \\ &- \sum_{v\in F_3/E} (A_v^{(1)} + A_v^{(2)}) - \sum_{e\in F_1} Z_e^{(1)} - \sum_{e\in F_2} Z_e^{(2)} - \sum_{e\in F_3} Z_e^{(3)} \\
    & - \sum_{v\in F_4/E,F_5/E,F_6/E} (A_v^{(1)}A_v^{(2)}+ A_v^{(2)}A_v^{(3)})\cr & - \sum_{e\in F_4,F_5,F_6} Z_e^{(1)}Z_e^{(2)}Z_e^{(3)}  - \sum_{v\in E_{12}/V} A_v^{(3)} - \sum_{v\in E_{23}/V} A_v^{(1)}\\& - \sum_{v\in E_{31}/V} A_v^{(2)}- \sum_{\substack{v\in E_{jk}/V \\ 4\le j<k\le 6}}(A_v^{(1)}A_v^{(2)}+ A_v^{(2)}A_v^{(3)})\\
    & - \sum_{v\in E_{14}/V,E_{15}/V,E_{16}/V} A_v^{(2)}A_v^{(3)} - \sum_{v\in E_{24}/V,E_{25}/V,E_{26}/V} A_v^{(3)}A_v^{(1)} \\ &- \sum_{v\in E_{34}/V,E_{35}/V,E_{36}/V} A_v^{(1)}A_v^{(2)} \\
    & - (A_{v_{456}}^{(1)}A_{v_{456}}^{(2)}+ A_{v_{456}}^{(2)}A_{v_{456}}^{(3)}) - A_{v_{145}}^{(2)}A_{v_{145}}^{(3)}\cr &  - A_{v_{246}}^{(3)}A_{v_{246}}^{(1)} - A_{v_{356}}^{(1)}A_{v_{356}}^{(2)}~,
    \end{split}
\end{align}
\end{small}
where $F,E,V$ are set of faces, edges, and corner vertices of the cube. $F_j/E$ denotes the set of vertices contained in the face $F_j$ but not included in its boundary edges. Other notations $\text{cube}/F, E_{jk}/V$ are defined in a similar fashion. $v_{jkl}$ denotes a corner vertex where the faces $F_j,F_k,F_l$ meet. $A_v$ is the vertex $X$ stabilizer of homological codes, which is defined by truncation at the boundary. $B_p$ is the plaquette $Z$ stabilizer of the homological codes.

This homological code on a cube has a logical $T$ gate. It is implemented by the unitary
\begin{align}
    U = &\prod_{c\in \text{cube}}(-1)^{a_1\cup a_3\cup a_2|_c} \prod_{p\in F_{4}, F_{5}, F_{6}} (-i)^{ [a_1]\cup [a_2]|_p} \cr&\quad\qquad  \cdot\prod_{e\in E_{35},E_{36}} e^{\frac{2\pi i}{8} [a_1]|_e}~,
\end{align}
where the first term $(-1)^{a_1\cup a_3\cup a_2|_c}$ is the product of six CCZ gates on a single cube $c$. The second term $(-i)^{ [a_1]\cup [a_2]|_p}$ is the product of two CS$^\dagger$ gates on a single plaquette $p$. The third term $e^{\frac{2\pi i}{8} [a_1]|_e}$ is the T gate on a single edge $e$. This unitary $U$ commutes with $X$ stabilizers within the Hilbert space that preserves the $Z$ stabilizers.

\section{More Logical Gates from Cohomology Operations}
\label{sec:moreexample}

There are more cohomology operations other than cup products reviewed in Sec.~\ref{sec:commutationCnZ} and the higher Pontryagin powers discussed in Sec.~\ref{sec:higherPontryagin}. In this section, we will comment on additional cohomology operations given by Steenrod square \cite{Steenrod:1947}, and use the operations to build new logical gates. For a review of Steenrod square and other cohomology operations, see e.g. \cite{mosher2008cohomology}.

The additional cohomology operations have two applications: (1) new logical gates from the SPT phases whose effective actions are given by the cohomology operation, as in Sec.~\ref{sec:higherPontryagin}; (2) In addition, the boundary of SPT phases can give new logical gates, as in Sec.~\ref{sec:logicalgatebdySPT}. We will focus on (1), and briefly comment on (2) with details to be published in a separate work.

\subsection{Properties of Steenrod square operations}

Steenrod square is a cohomology operation in $\mathbb{Z}_2$ coefficient, and it can be defined using higher cup product $\cup_i$ with $i\geq 1$ (when $i=0$ this is the ordinary cu product, and $\cup_0$ is written as $\cup$). 
We review the definition of higher cup product in appendix \ref{sec:higherP}. 
While higher cup product is not a cohomology operation, and in particular it is not associative, combination of higher cup product given by Steenrod square is a cohomology operation. 

\subsubsection{Definition of Steenrod Square operations}

The Steenrod square $Sq^i$ on $\mathbb{Z}_2$ $p$-cocycle $f$ is a $\mathbb{Z}_2$ $(p+i)$-cocycle $Sq^i f$, given by \cite{Steenrod:1947}
\begin{equation}
    Sq^if=f\cup_{p-i} f~.
\end{equation}
In particular, $Sq^0=1$, and $Sq^if=0$ for $p$-cochain $f$ such that $i>p$. 

Steenrod square operations satisfy the Cartan formula:
\begin{equation}\label{eqn:Cartan}
    Sq^i(f\cup g)=\sum_{n=0}^i Sq^n f\cup Sq^{i-n} g+d\zeta(f,g)~, 
\end{equation}
where $\zeta$ is called the Cartan coboundary and the expression can be found in \cite{medinamardones2019effectiveproofcartanformula} (see also \cite{Barkeshli:2021ypb,Barkeshli:2022edm} for examples).

Another important property of Steenrod square is that $Sq^1=d/2$ on $\mathbb{Z}_2$ cocycles, and it measures whether a $\mathbb{Z}_2$ cocycle can be lifted to a $\mathbb{Z}_4$ cocycle. For example, if $z$ is an integer cocycle, $dz=0$ not just mod $2$, then $Sq^1z=dz/2=0$ is trivial.

\subsection{Logical gates from Steenrod square operations}

As an example, consider 3-form $\mathbb{Z}_2$ homological code in 8D. Consider the logical gate
\begin{equation}
    (-1)^{\int C_3 Sq^2 C_3}=(-1)^{\int C_3\cup (C_3\cup_1 C_3)}~,
\end{equation}
where $(-1)^{C_3}=Z$ is the physical Pauli $Z$ gate on cubes (or 3-simplices), and it can be written as explicit physical gate on any space that admits a triangulation.

For example, if the space has the topology of $\mathbb{RP}^8$, where the cohomology is a ring generated by a 1-form $x$, there is a single logical qubit $n=0,1$ where $C_3=n x^3$. Using the Cartan formula (\ref{eqn:Cartan}) for $i=2,f=x^2,g=x$ and $Sq^2x=0$ since $2>1$, we find $Sq^2 C_3=nx^5$ and thus $\int C_3\cup Sq^2C_3=n\int x^8=n$. In other words, the logical gate on $\mathbb{RP}^8$ topolgy space is simply the logical Pauli $Z$ gate. This also confirms that $C_3\cup Sq^2C_3$ is a nontrivial element in the group cohomology.

As another example, if the space has topology $T^3\times\mathbb{RP}^5$, let us denote the generator of cohomology of $T^3$ by $y_1,y_2,y_3$ for the 3 circles, and $x$ for $\mathbb{RP}^5$.
Then there are 8 logical qubits, and the gauge field $C_3$ can be parametrized as
\begin{align}
    C_3&=n_1x^3+n_2 y_1y_2y_3+n_3 xy_1y_2+n_4 xy_1y_3+n_5xy_2y_3\cr &\quad +n_6 x^2y_1+n_7 x^2y_2+n_8 x^2y_3~,
\end{align}
where $n_i=0,1$ label the logical qubits.
Using the Cartan formula, and $Sq^1 y_i=0$ since $y_i$ are integer cocycles (they are volume forms of circles in $T^3$), we find
\begin{equation}
    Sq^2 C_3=n_1 x^5+n_6x^4y_1+n_7x^4y_2+n_8x^4y_3~.
\end{equation}
The integral $\int C_3\cup Sq^2C_3$ over $T^3\times \mathbb{RP}^5$ picks up the term $\int x^5y_1y_2y_3=1$ while all other terms integrates to zero, and thus 
\begin{align}
    (-1)^{\int C_3\cup Sq^2C_3}&=(-1)^{n_1n_2+n_5n_6+n_4n_7+n_3n_8}\cr &=\overline{CZ}_{1,2}\overline{CZ}_{5,6}\overline{CZ}_{4,7}\overline{CZ}_{3,8}~.
\end{align}
In other words, the operator on space of $T^3\times\mathbb{RP}^5$ produces product of fault-tolerant logical CZ gates.

\subsection{Generalization of folding in higher-form homological codes from SPT boundary}

Let us generalize the discussion in Sec.~\ref{sec:logicalgatebdySPT} to higher-form homological codes, described by higher-form gauge fields such as all loop homological code in 4 space dimensions and color code in 6 space dimensions. In such cases, the boundary map has additional choice of cocycle.

For subgroup $K$ of the bulk higher $m$-form gauge group $G$, their gauge fields satisfy
\begin{equation}
    dB_K=\eta(B_H)~,
\end{equation}
where $B_H,B_K$ are the Abelian higher $m$-form gauge fields for $H=G/K,K$, respectively, and $\eta$ describes the central extension of Abelian groups $K\rightarrow G\rightarrow H=G/K$.
On the boundary, $B_H=d\lambda_H$ is exact, with $(m-1)$-form gauge field $\lambda_H$. Since $\eta(d\lambda_H)$ is exact, we can redefine $B_K$ such that $dB_K=0$.
The boundary map ${\cal B}$ on $(m-1)$-form SPT $\omega_n\in H^n(B^{m}G,U(1)$ is given by
\begin{itemize}
    \item Choice of subgroup $K\subset G$ that is unbroken, such that $\omega_n|_K=d\alpha$. Here, $\alpha$ is a boundary action that depends on both $B_K$ and $\lambda_H$.

    \item Choice of cocycle for $B_K$ and $\lambda_H$, given by $\omega_{n-1}\in H^{n-1}(B^{m}K\times B^{m-1}H,U(1))$.
\end{itemize}
The boundary map ${\cal B}^{K,\omega_{n-1}}$ maps the higher-form SPT $\omega_n$ to $\alpha \omega_{n-1}$ (with group multiplication given by product in $U(1)$).

To understand the extra boundary action for $\lambda_H$,  consider 3D homological code, which can be expressed as 1-form homological code or 2-form homological code by exchanging the lattice with the dual lattice. The 1-form homological code admits 3 types of gapped boundaries 
\cite{Zhao:2022yaw,Ji:2022iva,PhysRevB.107.125425}: $e$ condensed, $m$ condensed, and twisted $m$ condensed-- where $m$ condensed boundary describes 2+1d $\mathbb{Z}_2$ gauge theory, and the twisted $m$ condensed boundary describes 2+1d $\mathbb{Z}_2$ gauge theory with nontrivial Dijkgraaf-Witten twist.
In terms of the 2-form homological code description, these gapped boundaries are the $m$ condensed boundary, $e$ condensed boundary and twisted $e$ condensed boundary.
The twisted $e$ condensed boundary differs from the $e$ condensed boundary by the Levin-Gu twist $(-1)^{\int \lambda_H^3}$ on the boundary.

\subsubsection{Example of boundary operation ${\cal B}$: $G=\mathbb{Z}_2^3$ 2-form homological code}

Consider $\mathbb{Z}_2^3$ 2-form homological code in 6D, which also describes the 6D color code \cite{Hsin:2024pdi}.
Let us denote the 2-form gauge fields by $a_1,a_2,a_3$, here the subscripts label different species. 

We begin with the cocycle $a_1\cup a_2\cup a_3+Sq^2(a_1\cup a_2)\in H^6(B^2\mathbb{Z}_2^3,U(1))$, which gives the SPT operator  $(-1)^{\int (a_1\cup a_2\cup a_3+Sq^2(a_1\cup a_2))}$, and apply the boundary map. 
Here, we can consider the boundary condition
\begin{equation}
    a_1+a_2+a_3=d\lambda~.
\end{equation}

To see this is a well-defined boundary condition, we note that the SPT cocycle is trivialized using the
Cartan formula
\begin{equation}
 Sq^2(a_1\cup a_2)=a_1^2 a_2+a_1a_2^2+\left(da_1/2\right)\left(da_2/2\right)+d\zeta(a_1,a_2)~, 
\end{equation}
where we have used $Sq^1a=da/2$, and
the Cartan coboundary $\zeta$ is a 5-cochain given in equation (143) of \cite{Barkeshli:2022edm}: on the 5-simplex (012345) it takes value
\begin{equation}
    \zeta(012345)=a_1(023)a_1(012)a_2(345)a_2(235)~.
\end{equation}
We note that $\zeta(a_1,a_2)$ is a product of two $a_1$ and two $a_2$.

Thus with $a_3=a_1+a_2$ up to an exact cocycle, $a_1a_2a_3+Sq^2(a_1a_2)$ is an exact cocycle $d\alpha$.
The boundary map gives
\begin{align}
    {\cal B}:\quad& (-1)^{a_1\cup a_2\cup a_3+Sq^2(a_1\cup a_2)} \cr &\mapsto \quad  (-1)^{a_1\cup a_2\cup \lambda}i^{[a_1]d[a_2]/2}(-1)^{\zeta(a_1,a_2)}
    ~.
\end{align}

The construction of logical gates using the boundary operation in higher-form homological codes is similar to the 1-form homological codes, and we will discuss it in a separate work.

 \section{Logical Gates on Good Quantum LDPC Codes from Cohomology Operation}
\label{sec:goodqLDPC}
\subsection{Building manifolds from quantum codes}
In this last section, we generalize the logical gates via cohomology operation to expander-based qLDPC codes including the good qLDPC code with constant rate and linear distance. 

Since the cohomology operation in this paper are mainly defined on a simplicial complex, which is not the case for general expander-based LDPC codes, we need a procedure to convert these qLDPC codes to a corresponding code defined on a simplicial complex.  In Ref.~\cite{freedman:2020_manifold_from_code}, a mapping for turning a quantum code based on a 3-term (2D) chain complex to a manifold with dimension 11 (or higher) has been developed. This mapping essentially erodes the distinction between general qLDPC (or even more generally CSS) codes and homological codes defined on manifolds. A homological code defined on a 11D triangulated manifold is produced from a general qLDPC code, and hence one can apply well-defined cohomology operation to implement the logical gates.   Nevertheless, so far this method only applies to qLDPC codes based on 3-term chain complex and hence we only discuss the application of logical Clifford gates in these codes.

We first briefly review the mapping in Ref.~\cite{freedman:2020_manifold_from_code}, more detailed and pedagogical introduction can be found in Refs.~\cite{zhu2025topological, guemard2025lifting}. Roughly speaking, the essential strategy is to subdivide the general chain complex of the quantum code into a simplicial complex (triangulation) of a manifold. First of all, for a large class of qLDPC codes or more generally a CSS codes corresponding to a 3-term (2D) $\ZZ_2$ chain complex, there exists a sparse lift from it to a chain complex over integer coefficients $\ZZ$,\footnote{
The sparseness is for the manifold to have bounded curvature, and thus the corresponding homological code on the triangulation is still LDPC \cite{freedman:2020_manifold_from_code}.
} since the underlying chain complex of a manifold is a $\ZZ$ chain complex. We say these code are sparsely liftable. The corresponding 3-term chain complex has the following form:
\begin{equation}\label{eq:lifted_complex}
     \tilde{C}_{2} \xrightarrow[]{\tilde{\partial}_{q+1}=\widetilde{\mathbf{H}}_Z^T} \tilde{C}_1 \xrightarrow[]{\tilde{\partial}_{q}=\widetilde{\mathbf{H}}_X} \tilde{C}_{0}~,
\end{equation}
where $\tilde{\partial}$ and $\widetilde{\mathbf{H}}$ are the lifted boundary map  and parity check matrices with $\ZZ$ coefficients respectively.   

Now there is an obvious obstruction for a naive mapping as follows: if the qubits are still placed on 1-cells (edges) of a  2D manifold, one is not able to map a general quantum code such as quantum expander codes to such a manifold since on a manifold a 1-cell (edge) can only be adjacent to two 0-cells (vertices), which then forbids the mapping for a general code where a qubit can be coupled to more than two $X$-checks. To circumvent this obstruction, we have to place the qubits to cells of at least two dimensions.   It was pointed out in Ref.~\cite{freedman:2020_manifold_from_code}, there may exist spurious 1-cycles and 2-cycles in the case where qubits are placed on 2-cells, which could shorten the code distance.  Therefore, it was proposed to place the qubits on 4-cells in order to have a separation between the dimension of the qubit and checks and the spurious cycle dimensions.  

The construction of the manifold then follows the procedure of handle decomposition.  The boundary maps in the lifted chain complex (Eq.~\eqref{eq:lifted_complex}), which contain the information about how the higher cells are attached to the lower cells, are translated to the recipe of how the higher dressed handles are attached to the lower dressed handles.  Here, the dressed handle is actually a handle-body which is itself composed of handles of the same and lower dimensions.  In short, the $X$-stabilizers, qubit, and $Z$-stabilizers will be mapped to a dressed 3-handle, dressed  4-handle, and dressed 5-handle respectively.  The details of the dressed handles can be found in Refs.~\cite{freedman:2020_manifold_from_code, zhu2025topological, guemard2025lifting}.  After attaching the dressed 3-, 4- and 5-handles to form a 5-handlebody, one obtains a manifold $M$ with boundary $\partial M$.   Now applying a doubling of the manifold $M$ according to Poincar\'e duality by attaching the two identical copies of $M$ together along the common boundary $\partial M$, one can obtain a closed 11-manifold $\mathcal{M}=D(M)$, where $D$ represents the doubling.  More generally, one can obtain an $(2j+3)$-manifold using such a construction if placing qubits on $j$-cells for $j \ge 4$.

Here, we focus on the construction of the one with minimal dimension, i.e., an 11-manifold.  This gives rise to a homological code defined on the triangulation $\mathcal{L}$ of the 11-manifold $\M$, which can be obtained from numerical algorithms.  This new code can be considered as a subdivided version of the input qLDPC code.  The corresponding chain complex is
\begin{align}
&C_{11} \rightarrow \cdots \rightarrow C_{5} \xrightarrow[]{\partial_{5}=\mathbf{H}_Z^T} C_4 \xrightarrow[]{\partial_{4}=\mathbf{H}_X} C_{3} \rightarrow \cdots \cr
&\qquad   \quad \ \ Z\text{-stabilizer}  \quad \   
 \text{qubit} \qquad X\text{-stabilizer}. \cr
\end{align}
Note that $Z$-stabilizers, qubits and $X$-stabilizers correspond to 5-cells, 4-cells and 3-cells respectively.   The logical-$Z$ operators correspond to 4-cycles on the triangulation $\mathcal{L}$ of the manifold $\M$. The logical-$X$ operators correspond to 4-cocycles or equivalently 7-cycles on the dual triangulation $\mathcal{L^*}$.  We hence call it a (4,7)-homological code.  Note that the code-to-manifold mapping in Ref.~\cite{freedman:2020_manifold_from_code} does not change the code parameter scaling including the encoding rate and code distance which now corresponds to the 4-systole and 7-systole of the manifold $\M$. The only changes of the code parameters are just certain constants.    For example, if the input code is a good qLDPC code \cite{pkldpc22, Quantum_tanner}, the output (4,7)-homological code will also be good, i.e., with constant encoding rate and linear distance, which has been used to construct a 3D local code with optimal parameters saturating the Bravyi-Poulin-Terhal bound \cite{portnoy2023local}. If the input code is a hypergraph product code \cite{Tillich:2014_hyergraph_product} of two good classical LDPC codes, then the corresponding (4,7)-homological code will have constant rate and $\sqrt{\mathbf{n}}$ distance ($\mathbf{n}$ represents the total qubit number)  \cite{freedman:2020_manifold_from_code}.

\subsection{Logical Clifford gates}
In the gauge theory language, this (4,7)-homological code corresponds to a higher-form (4-form) $\ZZ_2$ gauge theory described by the following action:
\begin{equation}
S=\frac{2}{2\pi}\int a^{4} db^{7}~,
\end{equation}
where $a^4$ and $b^7$ represent the electric and magnetic gauge fields respectively.   

One can alternatively define a dual (7,4)-homological code supported on the same manifold $\M$, where the qubits are put on the 7-cells (dual of 4-cells). It corresponds to the following 7-form gauge theory:
\begin{equation}
S'=\frac{2}{2\pi}\int a'^{7} db'^{4}~,
\end{equation}
where $a'^7$ and $b'^4$ represent the electric and magnetic gauge fields respectively.

The logical Clifford gates between two copies of codes, a (4,7)-homological code and a (7,4)-homological code,   can be described by the following cohomology operation:
\begin{align}
U=& (-1)^{\oint_{\M} a^{4} \cup a'^7},   
\end{align}
where $a^4$ and $a'^7$ are the operator-valued 4-cocyles and 7-cocycles of the two homological codes respectively.  Here, $U$ corresponds to a constant-depth local circuit composed of a product of interleaving CZ gates according to the cup-product rule.   Now we can again rewrite $U$ in terms of logical gates using the cohomology basis $\{\alpha^4\}$ and $\{\beta^4\}$ for each copy of code  ($a^4=\sum_{\alpha^4}\hat{m}_\alpha  \alpha^4$ and $a'^7=\sum_{\beta^7}\hat{m}'_\beta  \beta^7$) as
\begin{align}
U=& (-1)^{\sum_{\alpha^4,\beta^7 }\hat{m}_\alpha \hat{m}'_\beta \oint_{\M} \alpha^{4} \cup \beta^4} \cr
=&\prod_{\alpha^4,\beta^7 } \overline{\text{CZ}}(\alpha^4, \beta^7)^{\oint_\M \alpha^{4} \cup \beta^7} \cr
=& \prod_{\alpha^4,\beta^4 } \overline{\text{CZ}}(\alpha^4, \beta^7)^{|\alpha^*_7 \cap \beta^*_4|}~,
\end{align}
where  $|\alpha^*_7 \cap \beta^*_4|=\oint \alpha^{4} \cup \beta^7$ represents the intersection number of the dual cycles $\alpha^*_7$, $\beta^*_4$.  

Further extension of the cohomology operation to implement logical non-Clifford gates in 3D homological product codes can be found in Ref.~\cite{zhu2025topological}.

\vspace{1em}

\section{Outlook}
\label{sec:discussion}

Here are a few directions we will pursue in the future:
\begin{itemize}
    \item We can apply the current classification of logical gates in this paper to more general codes beyond the homological codes including expander-based qLDPC codes, as well as general CSS codes.

    \item We would like to understand better the algorithm-efficient logical gates such as $CR_k$ gates and how $k$ scales with the code parameters, and search for a much larger classes of new gates.

    \item We would like to understand better the logical gates in self-correcting quantum codes from the boundary operations, such as $CR_k$ gates in 2-form homological codes.

    \item Another direction is applying the method to non-Abelian codes such as $\mathbb{D}_8$ gauge theory and its generalizations \cite{Iqbal:2023wvm,Hsin:2024pdi}.
    
\end{itemize}

\section*{Acknowledgment}

We thank Ben Brown, Yu-An Chen, Ike Chuang, Michael Freedman, Anton Kapustin, Elia Portnoy and Zhenghan Wang for discussions. 
P.-S.H. is supported by Department of Mathematics King’s College
London. 
R.K. is supported by the U.S. Department of Energy through grant number DE-SC0009988 and the Sivian Fund.
G.Z. is supported by the U.S. Department of Energy, Office of Science, National Quantum
Information Science Research Centers, Co-design Center
for Quantum Advantage (C2QA) under contract number
DE-SC0012704.

\bibliography{biblio, mybib_merge.bib}{}

\appendix

\begin{widetext}

\section{Review of Cup Product on Lattices}
\label{sec:cupproductrev}

Let us review cup product for $\mathbb{Z}_2$ valued cochains on triangulated lattice and hypercubic lattices. For more details, see e.g. \cite{milnor1974characteristic,Benini:2018reh, PhysRevB.101.035101,Chen:2021ppt,Chen:2021xuc}.

A $\mathbb{Z}_2$-valued $m$-cochain $f_m$ is a map from $m$-simplices on the lattice to $0,1$ mod 2.
The cup product of $m$-cochain $f_m$ and $n$-cochain $g_n$ is an $(m+n)$-cochain $f_m\cup g_n$ defined as follows:
\begin{itemize}
    \item For triangulated lattice, denote $k$-simplices by the vertices $s_k=(0,1,2,3,\cdots, k)$, the cup product takes the following value on $(m+n)$-simplices:
    \begin{equation}
        f_m\cup g_n(0,1,2,\cdots, m+n)=
        f_m(0,1,\cdots,m)g_n(m,m+1,\cdots,m+n)~.
    \end{equation}
    We note that there is a common vertex $m$.

    \item For hypercubic lattice, the cup product $f_m\cup g_n$ on $(m+n)$-dimensional hypercube $s_{m+n}$ that span the coordinates $(x^1,x^2,\cdots x^{m+n})\in [0,1]^{m+n}$ is given by
    \begin{equation}
        f_m\cup g_n(s_{m+n})
        =\sum_I f_m\left([0,1]^I\right)g_n\left((x^I=1,x^{\bar I}=0)+[0,1]^{\bar I}\right)~,
    \end{equation}
    where the summation is over the different sets $I$ of $m$ coordinates out of the $(m+n)$ coordinates $x^1,\cdots x^{m+n}$, $\bar I$ denotes the remaining $n$ coordinates. See Figure \ref{fig:cup} for illustrations. 
    In each term in the sum, $f_m,g_n$ are evaluated on an $m$-dimensional hypercube and an $n$-dimensional hypercube, where the two hypercubes intersect at a point $(x^I=1,x^{\bar I}=0)$:
    \begin{itemize}
        \item $[0,1]^I$ is the $m$-dimensional unit hypercube starting from $(x^I=0,x^{\bar I}=0)$ and ending at $(x^I=1,x^{\bar I}=0)$, i.e. $[0,1]^I=\{0\leq x^i\leq 1,x^j=0:i\in I, j\in \bar I\}$.
        \item     $(x^I=1,x^{\bar I}=0)+[0,1]^{\bar I}$ is the $n$-dimensional hypercube in the $\bar I$ directions starting from $(x^I=1,x^{\bar I}=0)$ and ending at $(x^I=1,x^{\bar I}=1)$, i.e. $(x^I=1,x^{\bar I}=0)+[0,1]^{\bar I}=\{x^i=1,0\leq x^j\leq 1:i\in I,j\in \bar I\}$.
    \end{itemize}

\end{itemize}

\begin{figure}[t]
    \centering
    \includegraphics[width=\textwidth]{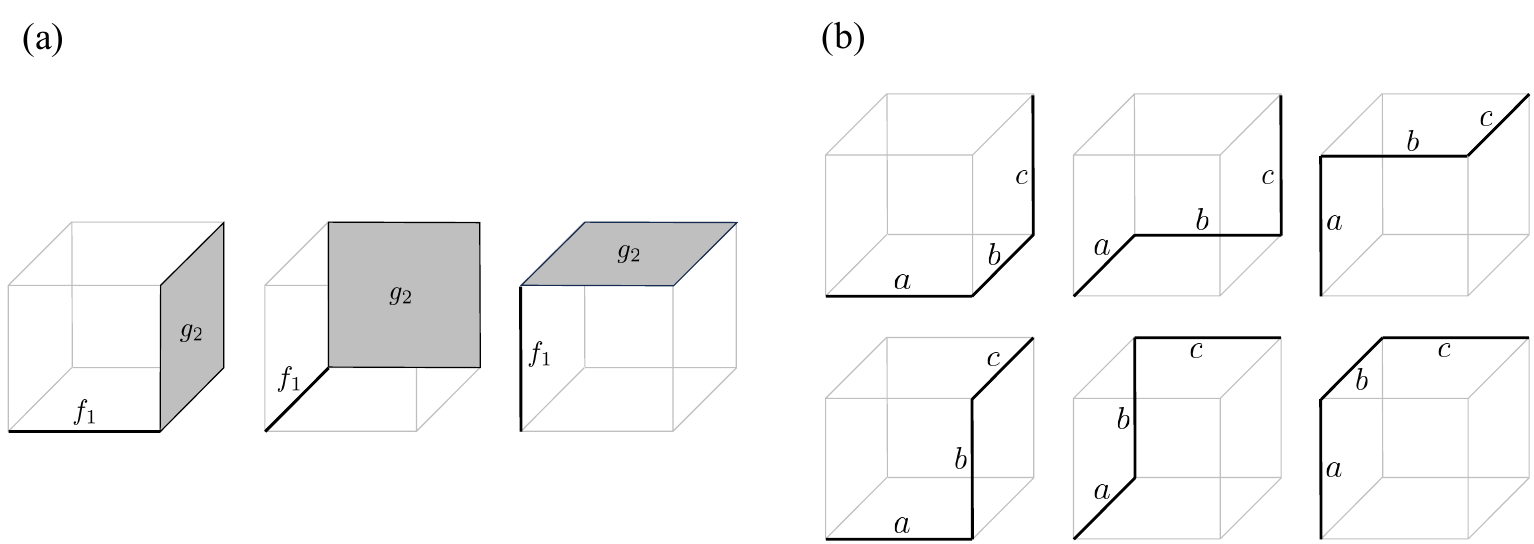}
    \caption{The cup product on a cube. (a): The cup product $f_1\cup g_2$ evaluated on a single 3d cube is the sum of these terms. (b): The cup product $a\cup b\cup c$ with 1-cochains $a,b,c$ evaluated on a single 3d cube is the sum of these terms.
    }
    \label{fig:cup}
\end{figure}

\section{Definition and Properties of Higher Pontryagin Powers}
\label{sec:higherP}

Let us define $n$th Pontryagin power for general $\mathbb{Z}_N$  $2r$-cocycles $a$ with $N=0$ mod $n$, where $n,r$ are integers. It gives a $2rn$-cocycle that takes value in $\mathbb{Z}_{nN}$:
\begin{equation}\label{eqn:higherpon}
    {\cal P}(a;n):=a^n+ \left(a\cup_1 da\right)a^{n-2}-\sum_{k=1}^{n-2}a\left(a^k\cup_1 da\right)a^{n-2-k}~,
\end{equation}
where we have omitted cup product $\cup$ since it is associative, while the higher cup products $\cup_i$ for $i\geq 1$ are not associative. We will review the definition of higher cup product below.
For $n=2$ and $r=1$ this is the Pontryagin square operation for even $N$ \cite{Whitehead1949}.

In the following, we will show that (1) ${\cal P}(a;n)$ is closed in $\mathbb{Z}_{nN}$, (2) ${\cal P}(a,n)$ mod $nN$ only depends on $a$ mod $N$, i.e. it does not depend on the lift of $a$ to integer cochain, and (3) ${\cal P}(a;n)$ only changes by an exact $\mathbb{Z}_{nN}$ cocycle under a ``gauge transformation'' $a\rightarrow a+d\chi$ for $\mathbb{Z}_{nN}$ $(2r-1)$-cochain $\chi$. Together they imply that ${\cal P}(\cdot;n)$ is a cohomology operation.

We remark that if $a$ can be lifted to a $\mathbb{Z}_{nN}$ cocycle, i.e. $da=0$ mod $nN$, then the Pontryagin power reduces to simply the cup product ${\cal P}(a;n)=a^n$.

\subsection{Definition of higher cup products}

For $p$ cochain $f$ and $q$ cochain $g$, their $\cup_i$ product produces a degree $(p+q-i)$ cochain $f\cup_i g$. Here we will provide the definition for $i=1$, while for higher $i$ the higher cup products can be found in \cite{Steenrod:1947}.
On $(p+q-1)$ simplex $(0,1,2,\cdots, (p+q-1))$, its value is defined to be
\begin{equation}
    f\cup_1 g(0,1,2,\cdots, (p+q-1)):=\sum_{j=0}^{p-1}(-1)^{(p-j)(q+1)}f(0,\cdots j,j+q,\cdots,p+q-1)g(j,\cdots,j+q)~.
\end{equation}
Higher cup products can also be defined on hypercubic lattice, see e.g. \cite{Chen:2021ppt} and Appendix C of \cite{Chen:2021xuc}.

\subsection{Useful higher cup product identities}
For convenience, here we collect a few identities for cup product $\cup$ and higher cup products $\cup_i$. For more details, see {\it e.g.} \cite{hatcher2002algebraic}, Appendix A of \cite{Benini:2018reh} and Appendix B of \cite{Barkeshli:2021ypb}: for degree $p,q,r$ cochains $f_p,g_q,h_r$,
\begin{align}
&f_p\cup g_q=(-1)^{pq}g_q\cup f_p+(-1)^{p+q+1}\left[
d(f_p\cup_1 g_q)-d f_p\cup_1 g_q - (-1)^p f_p\cup_1 d g_q
\right]\cr
&d(f_p\cup g_q)=d f_p\cup g_q+(-1)^p f_p\cup d g_q\cr
&d\left(f_p\cup_1 g_q\right)
=d f_p\cup_1 g_q+(-1)^p f_p\cup_1 d g_q + (-1)^{p+q+1} f_p\cup g_q+(-1)^{pq+p+q} g_q\cup f_p\cr 
&(f_p\cup g_q)\cup_1 h_r=(-1)^p f_p\cup (h_r\cup_1 g_q)+(-1)^{qr} (f_p\cup_1 h_r)\cup g_q~,
\end{align}
where the last identity is discovered in \cite{hirsch1955quelques}.

\subsection{Higher Pontryagin powers map between cocycles}

To see ${\cal P}(a;n)$ is closed in $\mathbb{Z}_{nN}$, we note that
\begin{align}
d(a^n)&=da a^{n-1}+ada a^{n-2}+\cdots + a^{n-1}da\cr 
&=\left(daa-ada\right)a^{n-2}+\sum_{k=1}^{n-2} a\left(a^k da-da a^k\right)a^{n-2-k}+nadaa^{n-2}\cr
&=\left(daa-ada\right)a^{n-2}+\sum_{k=1}^{n-2} a\left(a^k da-da a^k\right)a^{n-2-k}\text{ mod }nN~,
\end{align}
where the last line used $da=0$ mod $N$.
Further using the identity $a^\ell da-da a^\ell=d(a^\ell \cup_1 da)-d(a^\ell)\cup_1 da=d(a^\ell \cup_1 da)$ mod $N^2$, the equation becomes
\begin{align}
    d(a^n)&
= d\left(   -\left(a\cup_1 da\right)a^{n-2}+\sum_{k=1}^{n-2} a\left(a^k \cup_1 da\right)a^{n-2-k}\right)\text{ mod }nN~.
\end{align}
This implies ${\cal P}(a;n)$ defined in (\ref{eqn:higherpon}) is a $\mathbb{Z}_{nN}$ $2rn$-cocycle.

\subsection{Higher Pontryagin powers have well-defined values}

We also need to check ${\cal P}(a;n)$ mod $nN$ does not depend on the lift of $a$ to integer cochain. Suppose we change the lift of $a$ in $\mathbb{Z}_{nN}$ by $a\rightarrow a+ N x$ for integer $2r$-cochain $x$. Then ${\cal P}(a;n)$ changes by
\begin{align}
    &\frac{1}{N}\left( {\cal P}(a+Nx;n)-{\cal P}(a;n)\right)\cr 
    &=xa^{n-1}+axa^{n-2}+\cdots a^{n-1}x+\left(a\cup_1 dx\right)a^{n-2}
    -\sum_{k=1}^{n-2} a\left(a^k\cup_1 dx\right) a^{n-2-k}\text{ mod }N\cr 
    &=\left(xa-ax\right)a^{n-2}+\sum_{k=1}^{n-2}a\left(a^k x-xa^k\right)a^{n-2-k}
    +\left(a\cup_1 dx\right)a^{n-2}
    -\sum_{k=1}^{n-2} a\left(a^k\cup_1 dx\right) a^{n-2-k}\text{ mod }n~.\cr
\end{align}
Further using the identity $a^\ell x-x a^\ell=-d(a^\ell \cup_1 x)+d(a^\ell)\cup_1 x+a^\ell \cup_1 dx=-d(a^\ell \cup_1 x)+a^\ell \cup_1 dx$ mod $N$, the equation becomes
\begin{equation}
    \frac{1}{N}\left( {\cal P}(a+Nx;n)-{\cal P}(a;n)\right)=0\text{ mod }n~.
\end{equation}
In other words, ${\cal P}(a;n)$ has well-defined $\mathbb{Z}_{nN}$ value for $\mathbb{Z}_N$ cocycle $a$.

\subsection{Higher Pontryagin powers are gauge invariant}
Here we show the gauge invariance of the higher Pontryagin powers $\mathcal{P}(a;n)$, 
\begin{align}
    \mathcal{P}(a+d\chi;n) = \mathcal{P}(a;n) + d\phi~,
    \label{eq:gauge invaraince}
\end{align}
with some $(2rn-1)$-cochain $\phi$ in $\Z_{nN}$. Together with the invariance under $a\to a+Nx$, this implies that the higher Pontryagin power defines a cohomology operation
\begin{align}
    \mathcal{P}(\cdot;n): \quad H^{2r}(X, \Z_{N}) \to H^{2rn}(X, \Z_{nN})~.
\end{align}

To see \eqref{eq:gauge invaraince}, it is sufficient to check the invariance when $\chi$ is nonzero on a single $(2r-1)$-simplex, otherwise zero. General gauge transformation can be generated by combinations of such transformations.

Let us study the variation of $\mathcal{P}(a;n)$ on a $(2rn)$-simplex $(0,1,\dots,2rn-1)$ under the gauge transformation. We show that the gauge transformation by $a\to a+d\chi$ with $\chi$ nonzero on a single $(2r-1)$-simplex transforms
\begin{align}
    \mathcal{P}(a+d\chi;n)=\mathcal{P}(a;n) + d\phi(\chi,a)~,
\end{align}
with
\begin{align}
    \begin{split}
        \phi(\chi,a) &= \sum_{k=0}^{n-1} a^k\chi a^{n-k-1} + \sum_{k=0}^{n-2}(k+1)a^k(\chi\cup_1 da) a^{n-k-2} \\
         & \quad -\sum_{k+l+m=n-3 \atop k,l,m\ge 0} (m+1) a^{k}\chi a^{l} (a\cup_1da) a^{m} \\
        & \quad -\sum_{k+l+m=n-3 \atop k,l,m\ge 0} (l+m+1)a^{k}(a\cup_1da)a^{l}\chi a^{m} ~.\\
        \end{split} 
\end{align}

Below we check this by cases of a single $(2r-1)$-simplex in $(0,1,\dots,2rn-1)$ on which $\chi$ is nonzero.

\paragraph*{Case 1.}
    When $\chi$ is nonzero on a $(2r-1)$-simplex in $\partial(0,1,\dots 2r)$. In that case, the variation evaluated at the $(2rn)$-simplex $(0,1,\dots 2rn-1)$ is given by
    \begin{align}
        \begin{split}
            &\mathcal{P}(a+d\chi;n)-\mathcal{P}(a;n)\\
            &= d\chi a^{n-1} + (d\chi\cup_1 da) a^{n-2} -\sum_{k=1}^{n-2} d\chi (a^k\cup_1 da) a^{n-2-k} \\
            &= d\chi a^{n-1} + (\chi da +da\chi) a^{n-2} -\sum_{k=1}^{n-2} d\chi (a^k\cup_1 da) a^{n-2-k} + d((\chi\cup_1 da) a^{n-2}) \\
            &= d\chi a^{n-1} + (\chi da +da\chi) a^{n-2} -\sum_{k-1}^{n-2} \chi d(a^k\cup_1 da) a^{n-2-k} \\
            &\quad + d\left((\chi\cup_1 da) a^{n-2} - \sum_{k=1}^{n-2}\chi(a^k\cup_1 da)a^{n-2-k}\right) \\
             &= d\chi a^{n-1} + (\chi da +da\chi) a^{n-2} -\sum_{k-1}^{n-2} \chi (a^kda - da a^k) a^{n-2-k} \\
            &\quad + d\left((\chi\cup_1 da) a^{n-2} - \sum_{k=1}^{n-2}\chi(a^k\cup_1 da)a^{n-2-k}\right)~,\\
        \end{split}
    \end{align}
    where in the second equality we used $(d\chi\cup_1 da) a^{n-2}= (\chi da +da\chi) a^{n-2} + d((\chi\cup_1 da) a^{n-2})$ mod $N^2$. Also, since $\chi$ is zero except for $\partial(0,\dots,2r)$, the term $da\chi a^{n-2}$ vanishes. Then the above expression simplifies into
    \begin{align}
        \begin{split}
             &\mathcal{P}(a+d\chi;n)-\mathcal{P}(a;n)\\
             &= d\chi a^{n-1} + \chi da a^{n-2} - \sum_{k=1}^{n-2} \chi a^k da a^{n-2-k} + (n-2)\chi da a^{n-2} \\
             &\quad + d\left((\chi\cup_1 da) a^{n-2} - \sum_{k=1}^{n-2}\chi(a^k\cup_1 da)a^{n-2-k}\right) \\
             &= d\chi a^{n-1} - \chi da a^{n-2} - \sum_{k=1}^{n-2} \chi a^k da a^{n-2-k} \\
             &\quad + d\left((\chi\cup_1 da) a^{n-2} - \sum_{k=1}^{n-2}\chi(a^k\cup_1 da)a^{n-2-k}\right) \\
             &=  d\left(\chi a^{n-1} + (\chi\cup_1 da) a^{n-2} - \sum_{k=1}^{n-2}\chi(a^k\cup_1 da)a^{n-2-k}\right) \\
             &=  d\left(\chi a^{n-1} + (\chi\cup_1 da) a^{n-2} - \sum_{l=0}^{n-3}(n-2-l)\chi a^{l}(a\cup_1 da)a^{n-3-l}\right) \\
             &= d\phi(\chi,a)~,
        \end{split}
    \end{align}
where in the last equality we used the cochain-level identity
    \begin{align}
        a^k \cup_1 da = \sum_{l=0}^{k-1}a^l (a\cup_1 da) a^{k-l-1}~.
        \label{eq:hirsch}
    \end{align}
    
\paragraph*{Case 2.}
  
    When $\chi$ is nonzero on a $(2r-1)$-simplex in $\partial(2r,2r+1,\dots 4r)$.
    In that case, it is convenient to employ an alternative expression of the Pontryagin power:
    \begin{align}
        \mathcal{P}(a;n) = a^n+ \left(a\cup_1 da\right)a^{n-2}-\sum_{l=0}^{n-3}(n-2-l)a^{l+1}\left(a\cup_1 da\right)a^{n-l-3}~.
    \end{align}
    This can be derived by plugging \eqref{eq:hirsch} into the original expression of $\mathcal{P}(a;n)$.
Based on this new expression, the variation under the gauge transformation is given by
\begin{align}
        &\mathcal{P}(a+d\chi;n)-\mathcal{P}(a;n)\cr &= ad\chi a^{n-2} + (d\chi\cup_1 da)a^{n-2} - (n-2) a (d\chi\cup_1 da)a^{n-3} - \sum_{l=1}^{n-3}(n-2-l) ad\chi a^{l-1}(a\cup_1 da)a^{n-l-3} \cr
        &= ad\chi a^{n-2} + (\chi da+da\chi)a^{n-2} - (n-2) a (\chi da+da\chi)a^{n-3}  - \sum_{l=1}^{n-3}(n-2-l) a\chi a^{l-1}(ada-daa)a^{n-l-3} \cr
        &\quad + d\left((\chi\cup_1 da) a^{n-2}-(n-2)a(\chi\cup_1da)a^{n-3}\right)  - \sum_{l=1}^{n-3}(n-2-l) d(a\chi a^{l-1}(a\cup_1 da)a^{n-l-3}) ~.
\end{align}
Note that in our case $\chi da a^{n-2}=0, ada\chi a^{n-3}=0$. We then have
\begin{align}
        &\mathcal{P}(a+d\chi;n)-\mathcal{P}(a;n) \cr &=  ad\chi a^{n-2} + da\chi a^{n-2} - \sum_{l=0}^{n-3}(n-2-l) a\chi a^{l-1}(ada-daa)a^{n-l-3} \cr
        &\quad + d\left((\chi\cup_1 da) a^{n-2}-(n-2)a(\chi\cup_1da)a^{n-3}\right)  - \sum_{l=1}^{n-3}(n-2-l) d(a\chi a^{l-1}(a\cup_1 da)a^{n-l-3}) \cr
         &= ad\chi a^{n-2} + da\chi a^{n-2} - \sum_{l=0}^{n-3}a\chi a^{n-3-l}da a^{l}  \cr
         &\quad + d\left((\chi\cup_1 da) a^{n-2}-(n-2)a(\chi\cup_1da)a^{n-3}\right)  - \sum_{l=1}^{n-3}(n-2-l) d(a\chi a^{l-1}(a\cup_1 da)a^{n-l-3}) \cr
          &= d\left(a\chi a^{n-2} + (\chi\cup_1 da) a^{n-2}+2a(\chi\cup_1da)a^{n-3}\right)  - \sum_{l=1}^{n-3}(n-2-l) d(a\chi a^{l-1}(a\cup_1 da)a^{n-l-3}) \cr
         &= d\phi(\chi,a)~.
\end{align}

\paragraph*{Case 3.}    
    When $\chi$ is nonzero on a $(2r-1)$-simplex in $\partial(2mr,2mr+1,\dots 2mr)$ with $m\ge 3$.
\begin{align}
    \begin{split}
        &\mathcal{P}(a+d\chi;n)-\mathcal{P}(a;n)\\ 
        &= a^{m-1}d\chi a^{n-m} + (a\cup_1 da) a^{m-3} d\chi a^{n-m}  - \sum_{l=m-1}^{n-3}(n-2-l) a^{m-1}d\chi a^{l+1-m} (a\cup_1 da) a^{n-l-3} \\
        &\quad - \sum_{l=0}^{m-4} (n-2-l) a^{l+1}(a\cup_1 da)a^{m-l-4} d\chi a^{n-m} \\
        & \quad- (n-m+1)a^{m-2}(d\chi\cup_1da) a^{n-m} -(n-m) a^{m-1}(d\chi\cup_1da) a^{n-m-1} \\
        &= a^{m-1}d\chi a^{n-m} - (ada-daa) a^{m-3} d\chi a^{n-m} \\
        &\quad - \sum_{l=m-1}^{n-3}(n-2-l) a^{m-1}\chi a^{l+1-m} (ada-daa) a^{n-l-3} \\
        &\quad + \sum_{l=0}^{m-4} (n-2-l) a^{l+1}(ada-daa)a^{m-l-4} \chi a^{n-m} \\
        & \quad- (n-m+1)a^{m-2}(\chi da+da\chi) a^{n-m} -(n-m) a^{m-1}(\chi da + da\chi) a^{n-m-1} \\ 
        &\quad + d((a\cup_1da) a^{m-3}\chi a^{n-m}) \\
        & \quad - \sum_{l=m-1}^{n-3} (n-2-l) d(a^{m-1}\chi a^{l+1-m} (a\cup_1da) a^{n-l-3}) \\
        & \quad -\sum_{l=0}^{m-4} (n-2-l)d(a^{l+1}(a\cup_1da)a^{m-l-4}\chi a^{n-m}) \\
         & \quad- (n-m+1)d(a^{m-2}(\chi\cup_1da) a^{n-m}) -(n-m) d(a^{m-1}(\chi\cup_1da) a^{n-m-1}) ~.\\
    \end{split}
\end{align}
Note that in our case $a^{m-2}\chi da=0, a^{m-1}da\chi=0$. In the last equation, one can check that the terms that are not total derivative can be compiled into $d(a^{m-1}\chi a^{n-m})$ mod $N^2$. We then get
\begin{align}
    \begin{split}
     &\mathcal{P}(a+d\chi;n)-\mathcal{P}(a;n) \\ &= d(a^{m-1}\chi a^{n-m})  - \sum_{l=m-1}^{n-3} (n-2-l) d(a^{m-1}\chi a^{l+1-m} (a\cup_1da) a^{n-l-3}) \\
        & \quad -\sum_{l=0}^{m-3} (n-2-l)d(a^{l}(a\cup_1da)a^{m-l-3}\chi a^{n-m}) \\
         & \quad+ (m-1)d(a^{m-2}(\chi\cup_1da) a^{n-m}) +m d(a^{m-1}(\chi\cup_1da) a^{n-m-1}) \\
         &= d\phi(\chi,a)~.
    \end{split}
\end{align}

\paragraph*{Case 4.}

    When $\chi$ is nonzero on a $(2r-1)$-simplex which is not listed above, i.e., $d\chi$ vanishes on $2r$-simplex $(2mr,2mr+1,\dots,2(m+1)r)$ for any integer $m$. In that case, the variation under $a\to a+d\chi$ is given by
    \begin{align}
    \begin{split}
     \mathcal{P}(a+d\chi;n)-\mathcal{P}(a;n) &= (d\chi\cup_1 da)a^{n-2} - \sum_{l=0}^{n-3} (n-2-l) a^{l+1} (d\chi\cup_1 da) a^{n-l-3}~.
    \end{split}
\end{align}
Note that only the variations involving $\cup_1$ contributes to the expression, since the terms in the form of $(\dots\cup\chi\cup\dots)$ vanishes in our case. It can be written as
\begin{align}
    \begin{split}
         &\mathcal{P}(a+d\chi;n)-\mathcal{P}(a;n)= \sum_{l=0}^{n-2}(l+1)a^l (d\chi\cup_1 da)a^{n-2-l}= \sum_{l=0}^{n-2}(l+1)d(a^l(\chi\cup_1 da)a^{n-2-l}) \\
         &= d\phi(\chi,a)~.
    \end{split}
\end{align}

\subsection{Intersection properties}
\label{sec:refinecupproduct}

Let us show the higher Pontryagin powers are refinement of powers of cup products. This generalizes the quadratic refinement of Pontryagin square.

To show this, consider $\mathbb{Z}_N$ $2r$-cocycles $a,a'$:
\begin{equation}
    N \int\left( {\cal P}(a+a';n)-{\cal P}(a;n)-{\cal P}(a';n)\right)\text{ mod }nN~.
\end{equation}
Since $N$ is a multiple of $n$, the expression in the bracket only depends on its value mod $N$, and we can replace the higher Pontryagin powers by the cup product. Thus it is the same as
\begin{equation}
    N \int\left((a+a')^n-a^n-(a')^n\right)\text{ mod }nN
    =\sum_{k=1}^{n-1} N \left(\begin{array}{c}
          n\\k
    \end{array}\right)\int a^k \cup (a')^{n-k}\text{ mod }nN~.
\end{equation}
When $n$ is a prime number, the right hand side is always 0 mod $nN$, and thus the mixed terms have order $N$ and they are given by the ordinary cup product.

\section{Review of Condensation in Pauli Stabilizer Codes}
\label{sec:condesationZN}

In this appendix we review the condensation in $\mathbb{Z}_N$ $k$-form homological code. 
If we start with $\mathbb{Z}_N$ homological code, and choose a divisor of $N$ $\ell|N$, we can condense the electric charge $q_e=\ell$ to obtain $\mathbb{Z}_{\ell}$ homological code. To illustrate the construction, we start with the $k$-form homological code Hamiltonian
\begin{equation}
    H_0=-\sum_{s_{k-1}} \prod_{s_{k}:\partial s_{k}\supset s_
    {k-1}} X^{o(s_{k})}_{s_{k}}-\sum_{s_{k+1}}\prod_{s_k\in\partial s_{k+1}} Z^{o(s_k)}_{s_k}+\text{h.c.}~.
\end{equation}
We measure the $Z_{s_k}^\ell$ operators that create the electric charge $\ell$. The new stabilizers consist of the previous ones that commute with $Z_{s_k}^\ell$, and also $Z_{s_k}^\ell$:\footnote{
Examples of the procedure are discussed in e.g. \cite{Ellison:2021vth,Hastings:2021ptn,Hsin:2023ooo}.
}
\begin{equation}
    H=-\sum_{s_{k-1}} \prod_{s_{k}:\partial s_{k}\supset s_
    {k-1}} \left(X^{o(s_{k})}_{s_{k}}\right)^{N/\ell}-\sum_{s_{k+1}}\prod_{s_k\in\partial s_{k+1}} Z^{o(s_k)}_{s_k}-\sum_{s_k}Z_{s_k}^\ell+\text{h.c.}~.
\end{equation}
The ground states have $Z_{s_k}^\ell=1$ and thus we can restrict the $\mathbb{Z}_N$ qudits to be $\mathbb{Z}_\ell$ qudits, and the new theory is equivalent to $\mathbb{Z}_\ell$ $k$-form homological code.

\end{widetext}

\end{document}